\documentclass[aps,prd,showpacs, floatfix, letterpaper,preprintnumbers,superscriptaddress,nofootinbib,groupedaddress]{revtex4}

\pdfoutput=1
 \usepackage{dcolumn}
\usepackage{bm}
\usepackage{amsfonts}
\usepackage{relsize}
\usepackage{hyperref}

\usepackage[utf8]{inputenc}
\usepackage[font=scriptsize]{caption}
\usepackage{ulem}
\usepackage{amsmath}
\usepackage{amsfonts}
\usepackage{amssymb}
\usepackage{mathtools}
\usepackage{bm}
\usepackage{xcolor}
\usepackage{graphicx}
\usepackage{subfig}

\usepackage{soul}
\usepackage{titlesec}

\newcommand{\be}{\begin{equation}}
\newcommand{\ee}{\end{equation}}
\newcommand{\bea}{\begin{eqnarray}}
\newcommand{\eea}{\end{eqnarray}}

\DeclareMathAlphabet{\mathcal}{OMS}{cmsy}{m}{n}

\begin{document}

\title{Screening fifth forces in scalar-vector-tensor theories}

\author{Manuel Gonzalez-Espinoza}
\email{manuel.gonzalez@upla.cl}
\affiliation{Laboratorio de Did\'actica de la  F\'isica, Departamento de Matem\'atica, F\'isica y Computaci\'on, Facultad de Ciencias Naturales y Exactas, Universidad de Playa Ancha, Subida Leopoldo Carvallo 270, Valpara\'iso, Chile}
\affiliation{Laboratorio de investigación de Cómputo de Física, Facultad de Ciencias
Naturales y Exactas, Universidad de Playa Ancha, Subida Leopoldo Carvallo
270, Valparaíso, Chile}

\author{Giovanni Otalora}
\email{giovanni.otalora@academicos.uta.cl}
\affiliation{Departamento de F\'isica, Facultad de Ciencias, Universidad de Tarapac\'a, Casilla 7-D, Arica, Chile}

\author{Lucila Kraiselburd}
\email{lkrai@fcaglp.unlp.edu.ar}
\affiliation{Facultad de Ciencias Astron\'{o}micas y Geof\'{\i}sicas, Universidad Nacional de La Plata,\\
        Paseo del Bosque S/N, CP 1900, La Plata, 
        Argentina}
\affiliation{CONICET, Godoy Cruz 2290, CP 1425, Ciudad Aut\'{o}noma de Buenos Aires, Argentina}

\author{Susana J. Landau}
\email{slandau@df.uba.ar}
\affiliation{CONICET - Universidad de Buenos Aires, Instituto de Física de Buenos Aires (IFIBA) \\
        Ciudad Universitaria - Pab. I, 
        CP 1428, Ciudad Aut\'{o}noma de Buenos Aires,
        Argentina}

\date{\today}

\begin{abstract} 
We study a screening mechanism in the context of scalar-vector-tensor (SVT) theories. This screening mechanism is based on both the derivative self-interactions of the vector field and the interactions of the scalar field with the vector field and curvature. We calculate the field equations in a spherically symmetric space-time, and then, we study the conditions for which this mechanism is successful in a weak gravitational background. In order to corroborate these analytical results, we have performed a numerical integration of the full equations. 
Finally, the corrections to the gravitational potentials have also been computed. We conclude that the present model, including both kinds of interactions, can avoid the propagation of the additional longitudinal mode arising in these theories. We also show that the space parameter of  the model is compatible with solar system constraints. This result extends the previous one found in the literature for generalized Proca theories to the case of SVT theories in the presence of scalar-vector interactions. 
\end{abstract}


\maketitle



\section{Introduction}\label{Introduction}

One of the most puzzling challenges in cosmology is to explain the current accelerated expansion of the universe \cite{Riess:1998cb,Perlmutter:1998np}. According to the standard cosmological model ($\Lambda$CDM), a cosmological constant added to Einstein's equations produces the late-time acceleration of the universe and constitutes $68 \%$ of the total energy density along with the other $32\%$ associated with dark and ordinary matter \cite{Zwicky:1933gu}. On the other hand, it has been discussed that this proposal has theoretical problems such as the severe fine-tuning problem related to its energy scale, the so-called cosmological constant problem \cite{Weinberg:1988cp,Carroll:2000fy,Padilla:2015aaa}. 

Recently some tensions with increasing statistical significance have been found between estimations of cosmological observables that involve the assumption of the $\Lambda$CDM model and  values obtained from independent local measurements \cite{Abdalla:2022yfr,DiValentino:2020zio,DiValentino:2020vvd,Heisenberg:2022lob}. For instance, the value of the Hubble constant today $H_{0}$, inferred from Cosmic Microwave data provided by the Planck collaboration and assuming the $\Lambda$CDM model \cite{Planckcosmo2018}, is $4.\sigma$ to  $6.3 \sigma$ below  local estimations such as the one obtained from type Ia supernovae and Cepheid data provided  by the SHOES collaboration \cite{Riess2022}. A similar issue arises with the clustering amplitude $S_8$: the value obtained from Planck data and using the theoretical predictions of the $\Lambda$CDM model is above that obtained from low-redshift observations \cite{Heymans:2020gsg,Nunes:2021ipq,Heisenberg:2022lob,Heisenberg:2022gqk}.

Several alternative theoretical constructions have been proposed in the literature to address the aforementioned cosmological constant problem, and some of them could also alleviate the so-called Hubble tension. Among them, we can mention: i) dynamical scalar fields minimally coupled to gravity or matter with an appropriate potential (usually known as quintessence)\cite{Wetterich:1987fm,Ratra:1987rm,Carroll:1998zi,Tsujikawa:2013fta} or  non-canonical kinetic energy (also known as k-essence) \cite{Chiba:1999ka,ArmendarizPicon:2000dh,ArmendarizPicon:2000ah}; ii) dynamical scalar fields coupled to curvature or torsion or matter \cite{2004PhRvD..70l3518B,2004PhRvL..93q1104K,2007PhRvD..76f4004H,PhysRevD.77.046009, Lopez:2021agu,Gonzalez-Espinoza:2021qnv,Gonzalez-Espinoza:2021mwr,Gonzalez-Espinoza:2020jss,Gonzalez-Espinoza:2020azh,Gonzalez-Espinoza:2019ajd,Otalora:2014aoa,Otalora:2013dsa,Otalora:2013tba,Gonzalez-Espinoza:2023whd}; iii) dynamical scalar fields with higher order derivatives in the action (also known as Galileons) \cite{Nicolis:2008in,Deffayet:2009wt,Baker:2017hug,Sakstein:2017xjx} and many others \cite{2012PhR...513....1C}. 

The Lagrangian of the covariant Galileon is constructed to keep the equations of motion at second order, while recovering the Galilean symmetry in the limit of Minkowski space-time \cite{Deffayet:2009wt}. Since the equations of motion are kept up to second order in time and spatial derivatives,  this theory can avoid Ostrogradski’s instability \cite{Ostrogradsky:1850fid}. Moreover, the most general scalar-tensor theories with second-order equations of motion were first found by Horndeski in $1974$ \cite{Horndeski:1974wa}. The Horndeski theory includes Brans-Dicke theory, minimally and non-minimally scalar field models, covariant Galileon, among others.

Scalar fields are not the only possibility to account for the present accelerated expansion of the universe; namely, vector fields have also been considered \cite{Armendariz-Picon:2004say,Koivisto:2008xf,Gomez:2020sfz,Gonzalez-Espinoza:2022hui,DeFelice:2016yws,Rodriguez-Benites:2023otm}. For instance, in the context of generalized Proca theories, a massive vector field breaking the $U(1)$ gauge symmetry is introduced. It has been shown that its time-dependent component can lead to an accelerated expansion of the Universe by exhibiting an asymptotic de Sitter attractor \cite{DeFelice:2016yws,DeFelice:2016uil,Nakamura:2019phn,DeFelice:2020icf,Cardona:2022lcz}. These generalized Proca theories are the most general vector-tensor theories that lead to second-order equations of motion. The first attempt to construct a general vector-tensor theory dates back to $1976$ when Horndeski also derived the most general action of an Abelian vector field with non-minimal coupling to gravity, which satisfies second-order equations of motion \cite{Horndeski:1976gi}. In order to find this action, he assumed that the vector field respects the gauge symmetry and that Maxwell equations are recovered in the flat space-time limit. Nevertheless, keeping the field equations at second order and dropping the $U(1)$ gauge invariance opens the opportunity to introduce non-trivial terms associated with  Galileon-type derivative self-interactions in the framework of generalized Proca theories \cite{Heisenberg:2014rta,Tasinato:2014eka,Tasinato:2014mia,Allys:2015sht,BeltranJimenez:2016rff}.

These two prominent classes of theories, the scalar Horndeski and generalized Proca theories, can be unified in the context of scalar-vector-tensor (SVT) theories with second-order field equations \cite{Heisenberg:2018acv,Heisenberg:2018mxx}. The SVT theories are usually classified into two groups depending on whether the $U(1)$ gauge symmetry is respected.  When the $U(1)$ symmetry is kept, the total propagating degrees of freedom (DOFs) are five: one scalar, two transverse vectors, two tensor polarizations. In the case of a broken $U(1)$ symmetry, there is a propagating longitudinal scalar mode in addition to the other five DOFs, which is consistent with a gravitational theory that includes a massive spin-1 field and a scalar field. Furthermore, in the presence of derivative interactions, such as those appearing in Galileon theories, it is natural to inquire about the existence of additional degrees of freedom due to these modifications in gravity. In Ref. \cite{Heisenberg:2014rta}, the authors proposed a generalized Proca theory that includes derivative interactions within a second-order action, constituting the vector model used to construct the scalar-vector theory addressed in this paper. Through an analysis based on the Hessian matrix, they demonstrated that only the three DOFs of the original Proca theory can propagate. It is crucial to highlight that, within these theories, the fourth DOF from the massive vector field in the off-shell configuration is unphysical, acting as a ghost degree of freedom similar to a Boulware-Deser ghost in massive gravity \cite{deRham:2014zqa}. In the current theory, it is systematically removed order by order through a system of constraints. These constraints are established by ensuring that the determinant of the corresponding Hessian matrix vanishes \cite{Heisenberg:2014rta}.

In the present paper, we  study the propagation of the longitudinal component of a vector field in a spherically symmetrical background, and its effects on the behavior of the gravitational potentials in a scalar-vector-tensor theory. It is important to stress that a theory that is expected to successfully explain the present accelerated expansion of the universe must also be consistent with local gravity constraints \cite{Will2014,will2018theory,Hohmann:2017qje,Gonzalez-Espinoza:2021nqd}. For instance, a screening mechanism of the longitudinal mode is usually required to lead to the suppression of the propagation of the fifth force on local scales. This is similar to the Vainshtein mechanism \cite{Vainshtein:1972sx} for scalar Galileons \cite{Burrage:2010rs,DeFelice:2011th,Kimura:2011dc,Kase:2013uja}. In this sense, the screening mechanism of the longitudinal scalar mode for vector Galileons in the presence of derivative self-interactions was studied in Ref. \cite{DeFelice:2016cri}. In particular, they found that due to the cubic-order  derivative self-interactions, the screening mechanism of the longitudinal scalar mode can be sufficiently efficient to keep the theory consistent with solar-system constraints. Therefore, here we intend to extend these latter results to the case of SVT theories. 
On the other hand, any valid theory of gravity is severely constrained by solar system tests. Therefore, we use current constraints  on the parameterized post-Newtonian (PPN) parameter $\gamma$ \cite{2003Natur.425..374B} to test the validity of the particular cases of SVT theories that we consider in this paper.

The plan of the paper is the following: In Section \ref{2}, we present the total action of the model and the field equations. In Section \ref{3}, we study the analytical solutions inside and outside a spherically symmetric compact object.  In Section \ref{Num_Res}, we corroborate our previous analytical results by numerically solving the field equations focusing on the case where the compact object is the Sun, which is relevant for the observational constraints. In Section \ref{subsecII}, we study the first-order corrections to the gravitational potentials. We also discuss the values of the free parameters of the model that are not ruled out by solar system tests. Finally, in Section \ref{conclusion_f}, we summarize the obtained results.

\section{Field equations of motion}{\label{2}

In this section, we introduce the expressions for the Lagrangian in SVT theories with broken $U(1)$ symmetry \cite{Heisenberg:2018acv,Heisenberg:2018mxx}.  In this Lagrangian, the vector field is minimally coupled to gravity, ensuring that the velocity of the tensor modes remains equal to the speed of light \cite{Heisenberg:2018mxx}. Conversely, the scalar field is non-minimally coupled to gravity. We focus  on the propagation of the longitudinal component of the vector field within a spherically symmetrical background. 


In order to write the general action of a massive vector field $A_{\mu}$ interacting with a scalar field $\phi$ in the curved spacetime, we define the variables 
\begin{equation}
    X_1 = - \dfrac{1}{2} \nabla_\mu \phi \nabla^\mu \phi, \ \ \ \ \ X_2 =  - \dfrac{1}{2} A^\mu \nabla_\mu \phi, \ \ \ \ \ X_3 =  - \dfrac{1}{2} A_\mu A^\mu.
\end{equation}
Furthermore, we introduce the effective metric
\be
\mathcal{G}_{\mu \nu}^{h_{n}}=h_{n 1}(\phi,X_{i})g_{\mu \nu}+h_{n 2}(\phi,X_{i})\nabla_{\mu}{\phi}\nabla_{\nu}{\phi}+h_{n 3}(\phi,X_{i})A_{\mu} A_{\nu}+h_{n 4}(\phi,X_{i})A_{\mu} \nabla_{\nu}{\phi},
\ee where $h_{n i}$ are functions of $\phi$ and $X_{i}$ with $i=1,2,3$.
For the vector field we also define the field strength $F_{\mu \nu}$, its dual $\tilde{F}^{\mu \nu}$ and the symmetric tensor $S_{\mu \nu}$ in the following form 
\be
F_{\mu \nu}=\nabla_{\mu}A_{\nu}-\nabla_{\nu}A_{\mu}, \:\:\: \tilde{F}_{\mu \nu}=\frac{1}{2} \mathcal{E}^{\mu \nu \alpha \beta} F_{\alpha \beta},\:\:\: S_{\mu \nu}=\nabla_{\mu}A_{\nu}+\nabla_{\nu}A_{\mu}.
\ee The covariant derivative operator $\nabla_{\mu}$ is the standard one associated to the Levi-Civita connection. In this way, we also have
\be 
F=-\frac{1}{4} F_{\mu \nu} F^{\mu \nu},\:\:\: Y_{1}=\nabla_{\mu}{\phi}\nabla_{\nu}{\phi}F^{\mu \alpha} F^{\nu}_{~\alpha},\:\:\: Y_{2}=\nabla_{\mu}{\phi} A_{\nu} F^{\mu \alpha} F^{\nu}_{~\alpha},\:\:\: Y_{3}=A_{\mu} A_{\nu} F^{\mu \alpha} F^{\nu}_{~\alpha},
\ee  which encodes the interactions arising from the pure vector modes.
Finally, the double dual Riemann tensor $L^{\mu \nu \alpha \beta}$, as well as the intrinsic vector interactions represented through the $2$-rank tensors $\mathcal{M}_{5}^{\mu \nu}$ and $\mathcal{N}_{5}^{\mu \nu}$, and the $4$-rank tensors $\mathcal{M}_{6}^{\mu \nu \alpha \beta}$ and $\mathcal{N}_{6}^{\mu \nu \alpha \beta}$ are defined by
\bea
&& L^{\mu \nu \alpha \beta}=\frac{1}{4}\mathcal{E}^{\mu\nu \rho \sigma}\mathcal{E}^{\alpha \beta \gamma \delta}R_{\rho \sigma\gamma \delta},\:\:\: 
\mathcal{M}_{5}^{\mu \nu}=
\mathcal{G}_{\rho \sigma}^{h_{5}}\tilde{F}^{\mu \rho}\tilde{F}^{\nu \sigma},\:\:\: \mathcal{N}_{6}
^{\mu \nu}=\mathcal{G}_{\rho \sigma}^{\tilde{h}_{5}}\tilde{F}^{\mu \rho}\tilde{F}^{\nu \sigma},\nonumber\\
&& \mathcal{M}_{6}^{\mu \nu \alpha \beta}=
2 f_{6,X_1}(\phi,X_1)\tilde{F}^{\mu \nu}\tilde{F}^{\alpha \beta},\:\:\: \mathcal{N}_{6}
^{\mu \nu \alpha \beta}=\frac{1}{2}\tilde{f}_{6,X_3}(\phi,X_{3})\tilde{F}^{\mu \nu}\tilde{F}^{\alpha \beta},
\eea where $\tilde{h}_{5j} (j=1,2,3,4)$ are also functions $\phi$, $X_{1}$, $X_{2}$ and $X_{3}$.

Therefore, the general action of SVT theories with broken $U(1)$ gauge symmetry is written as  \cite{Heisenberg:2018acv,Heisenberg:2018mxx}
\be
\label{SVT}
S_{SVT}=\int{d^{4}x\sqrt{-g}\sum_{n=2}^{6}{\mathcal{L}_{n}}},
\ee with the Lagrangians
{\small
\bea
&& \mathcal{L}_{2}=f_{2}(\phi,X_{1},X_{2},X_{3},F,Y_{1},Y_{2},Y_{3}), \nonumber\\
&& \mathcal{L}_{3}=f_{3}(\phi,X_{3}) g^{\mu\nu} S_{\mu \nu}+\tilde{f}_{3}(\phi,X_{3})A^{\mu}A^{\nu}S_{\mu \nu},\nonumber\\
&&\mathcal{L}_{4}=f_{4}(\phi,X_{3}) R+f_{4,X_{3}}(\phi,X_{3})\left[(\nabla_{\mu}{A^{\mu}})^{2}-\nabla_{\mu}{A_{\nu}}\nabla^{\nu}{A^{\mu}}\right],\nonumber\\
&& \mathcal{L}_{5}=f_{5}(\phi,X_{3}) G^{\mu \nu} \nabla_{\mu}{A_{\nu}}-\frac{f_{5,X_{3}}(\phi,X_{3})}{6}\left[(\nabla_{\mu}{A^{\mu}})^{3}-3\nabla_{\mu}{A^{\mu}}\nabla_{\rho}{A_{\sigma}}\nabla^{\sigma}{A^\rho}+2\nabla_{\rho}{A_{\sigma}}\nabla^{\gamma}{A^{\rho}}\nabla^{\sigma}{A_{\gamma}}\right]+\nonumber\\
&& \mathcal{M}_{5}^{\mu \nu} \nabla_{\mu}\nabla_{\nu}{\phi}+\mathcal{N}_{5}^{\mu \nu}S_{\mu \nu},\nonumber\\
&& \mathcal{L}_{6}=f_{6}(\phi,X_{1}) L^{\mu \nu \alpha \beta} F_{\mu \nu}F_{\alpha \beta}+\mathcal{M}_{6}^{\mu \nu \alpha \beta}\nabla_{\mu}\nabla_{\alpha}{\phi} \nabla_{\nu}{\nabla_{\beta}\phi}+
\tilde{f}_{6}(\phi,X_{3})L^{\mu \nu \alpha \beta} F_{\mu \nu}F_{\alpha \beta}+\mathcal{N}_{6}^{\mu \nu \alpha \beta} S_{\mu \alpha} S_{\nu \beta},
\eea} where $R$ and $G^{\mu  \nu}$ are the Ricci scalar and the Einstein tensor, respectively, and $f_{4,X_3}\equiv \partial f_{4}/\partial{X_3}$, $f_{5,X_3}\equiv \partial f_{5}/\partial{X_3}$. 
In order to obtain the full SVT action with second-order equations one could also add to \eqref{SVT} the action of scalar-tensor Horndeski theories. However, we focus only on the action \eqref{SVT},  since we are interested in the effects on the gravitational potentials due to the vector interactions. 

Interestingly enough, the action \eqref{SVT}  can be simplified using the results of recent observational data. For instance, for late-time cosmology, there is a tight bound on the speed of the tensor modes $c_{t}$ constrained from the gravitational event GW170817 \cite{LIGOScientific:2017vwq} and the gamma-ray burst GRB 170817A \cite{Goldstein:2017mmi}, which gives $-3\times 10^{-15} \leq c_{t}-1 \leq 7 \times 10^{-16}$. Thus, to guarantee $c_{t}=1$ one should assume $f_{4}(\phi,X_{3})=f_{4}(\phi)$ and $f_{5}(\phi,X_{3})=const$ \cite{Heisenberg:2018mxx}. In this case, the Lagrangian $\mathcal{L}_{4}$ only contributes to the field equations through the first term, while the Lagrangian $\mathcal{L}_{5}$ contributes through the third and fourth term. Furthermore, we are interested in studying the local gravity constraints for a viable dark energy model, which is well described by the flat Friedmann-Lemaître-Robertson-Walker (FLRW) metric  \cite{Copeland:1997et}. It is well known that the quantities $F$, $Y_1$, $Y_2$, $Y_3$ vanish on  this cosmological background  \cite{Heisenberg:2018mxx}. Also, the Lagrangian $\mathcal{L}_{6}$, along with the interactions proportional to $\mathcal{M}_{5}^{\mu \nu}$ and $\mathcal{N}_{5}^{\mu \nu}$ in $\mathcal{L}_{5}$, do not affect the background cosmology either \cite{Heisenberg:2018mxx}. Therefore, we can also neglect all these interactions in action \eqref{SVT}.

Thus, the relevant action that we consider in this work is given by
\begin{eqnarray}
&& S=\int d^4 x \sqrt{-g} \Big[ f_2(\phi,X_1,X_2,X_3,F) + f_3(\phi, X_3) g^{\mu\nu} S_{\mu\nu} + f_4(\phi) R \Big]+S_{m}(\Psi_{m},g_{\mu \nu}),
\label{action_final}
\end{eqnarray} where $S_{m}$ is action of the matter fields. 

Let us consider a spherically symmetric and static background whose line element is written as
\be
ds^{2}=-e^{2\Psi(r)}dt^{2}+e^{2\Phi(r)}dr^{2}
+r^{2} \left( d\theta^{2}+\sin^{2}\theta\,
d\varphi^{2} \right)\,,
\label{line}
\ee being $\Psi(r)$ and $\Phi(r)$ the gravitational potentials which are functions of the radius $r$.

The vector field can be expressed as 
\be
A^{\mu}=(\phi_A, A^{i}), 
\ee with $i=1,2,3$. By using Helmholtz’s theorem, the spatial components $A^{i}$ can be decomposed into the transverse and longitudinal modes as follows
\be
A_{i}=A_{i}^{(T)}+\nabla_{i}{\chi},  
\ee where $\chi$ is the longitudinal scalar and the transverse mode $A_{i}^{(T)}$ satisfies the traceless condition $\nabla^{i}{A_{i}^{(T)}}=0$. In a spherically symmetry background, the components $A_{\theta }^{(T)}$ and  $A_{\varphi}^{(T)}$ vanish. Furthermore, from the traceless condition, along with the regularity assumption for $A_{r}^{(T)}$ at $r=0$, one can show that the transverse vector $A_{i}^{(T)}$ also vanishes \cite{DeFelice:2016cri}. Therefore, we may focus only on the propagation of the longitudinal mode with the components of $A^{\mu}$ given by
\be
A^{\mu}=\left(\phi_A (r), e^{-2\Phi}\chi'(r),0,0 \right)\,.
\label{Amu}
\ee  
For the matter sector, we consider the perfect fluid with energy-momentum tensor 
\be
T_{\mu\nu}=\left[\left(\rho_{m}+P_{m}\right)U_{\mu}U_{\nu}+g_{\mu\nu}P_{m}\right],
\label{PFluid}
\ee with  $\rho_{m}$ and $p_{m}$ the energy density and pressure 
respectively. The 
four-velocity of the fluid is $U_{\mu}=(-e^{\Psi(r)},0,0,0)$ such that $U_{\mu} U^{\mu} 
=-1$ \cite{will2018theory}.

Thus, varying the action \eqref{action_final} with respect to the metric, we obtain the field equations for the latter:
\begin{eqnarray}
\mathcal{C}_1 \Psi'^2 + \mathcal{C}_2  \Psi'  + \left( \mathcal{C}_3 + \dfrac{\mathcal{C}_4}{r} \right) \Phi' +  \mathcal{C}_5 + \dfrac{\mathcal{C}_6}{r} + \dfrac{\mathcal{C}_7}{r^2} = - e^{2 \Phi } \rho_m ,\label{eom00} \\
\mathcal{C}_{8} \Psi'^2 + \left( \mathcal{C}_{9} + \dfrac{\mathcal{C}_{10}}{r} \right) \Psi' +  \mathcal{C}_{11} + \dfrac{\mathcal{C}_{12}}{r} + \dfrac{\mathcal{C}_{13}}{r^2} =  e^{2 \Phi } P_m ,\label{eom11}\\
\mathcal{C}_{14} \Psi'' + \mathcal{C}_{15} \Psi'^2  + \mathcal{C}_{16} \Psi' \Phi' + \left( \mathcal{C}_{17} + \dfrac{\mathcal{C}_{18}}{r} \right) \Psi' +  \mathcal{C}_{19} + \dfrac{\mathcal{C}_{20}}{r}  =  e^{2 \Phi } P_m ,\label{eom22}\\ \nonumber
\end{eqnarray}
 while the motion equations for the scalar and vector fields are obtained, as usual, varying the action with respect to each field
\begin{eqnarray}
\mathcal{D}_1 \Psi'' + \mathcal{D}_2 \Psi'' \Psi' + \mathcal{D}_3 \Psi'^2+  \mathcal{D}_4 \Psi'^3 + \mathcal{D}_5 \Phi'\Psi'^2  + \left( \mathcal{D}_6 + \dfrac{\mathcal{D}_7}{r} \right) \Psi'  &&\nonumber\\ 
 + \left( \mathcal{D}_8 + \dfrac{\mathcal{D}_9}{r} \right) \Phi' + \mathcal{D}_{10} \Phi'\Psi'
+  \mathcal{D}_{11} + \dfrac{\mathcal{D}_{12}}{r} + \dfrac{\mathcal{D}_{13}}{r^2} &=& 0 , \label{eomphi} \\
\mathcal{D}_{14} \Psi'' + \mathcal{D}_{15} \Psi'' \Psi' + \mathcal{D}_{16} \Psi'' \Psi'^2 + \mathcal{D}_{17} \Psi'^2+  \mathcal{D}_{18} \Psi'^3 +  \mathcal{D}_{19} \Psi'^4 + \mathcal{D}_{20} \Phi'\Psi'^2  &&\nonumber\\ 
 + \mathcal{D}_{21} \Phi'\Psi'^3  + \left( \mathcal{D}_{22} + \dfrac{\mathcal{D}_{23}}{r} \right) \Psi' + \mathcal{D}_{24} \Phi'+ \mathcal{D}_{25} \Phi'\Psi'
+  \mathcal{D}_{26} + \dfrac{\mathcal{D}_{27}}{r} + \dfrac{\mathcal{D}_{28}}{r^2} &=& 0 , \label{eomA0}\\
\mathcal{D}_{29} \Psi'
+  \mathcal{D}_{30} + \dfrac{\mathcal{D}_{31}}{r} + \dfrac{\mathcal{D}_{32}}{r^2} &=& 0 , \label{eomA1}
\end{eqnarray}
where coefficients $\mathcal{C}_{i}$ and $\mathcal{D}_{i}$ are shown in appendix \ref{Appendix0}.  Besides, to make our findings more accessible, you can follow the process of obtaining coefficients $\mathcal{C}_{i}$ and $\mathcal{D}_{i}$ in an online {\scshape Mathematica}\textsuperscript{\textregistered} notebook \cite{coefPPN}. \\


Below, for simplicity, we chose to analyze the following particular model,
\begin{equation}
    f_4 = \frac{M_{\rm pl}^2}{2} + \alpha_4 \phi^2, \ \ \ f_3 = \frac{1}{2} \beta_3 X_3 , \ \ \ f_2 = V(\phi) + X_1 + \beta_2 X_2 + m^2 X_3 + F, \label{particular_model}
\end{equation}
where $\alpha_4$ and $\beta_3$ are dimensionless constants, $m$ represents the vector field mass, and $\beta_2$ is a constant with the same dimension as $\phi$ and $\phi_A$, i.e., dimensions of mass. We aim to obtain analytical expressions of the gravitational potentials $\Phi(r)$ and $\Psi(r)$ under the weak field approximation.
\section{Analytical vector-scalar profiles}{\label{3}}
In order to  obtain approximate analytical solutions to the field equations, we divide the space of solutions according to two regions of interest: inside a spherically symmetric compact body of radius $r_{*}$, that is $r<r_*$,  and outside the body $r\geq r_*$. }
\\
\subsection{Solutions for \texorpdfstring{$r<r_*$}{Lg}}
Assuming a spherical density distribution $\rho_0$ of radius $r_*$ and GR potentials \eqref{GR_potential_inside} as leading order, the field equations within it are written as
\begin{eqnarray}
V'(\phi (r))+ \frac{2 \alpha _4 \rho _0 \phi (r)}{3 M_{\text{pl}}^2}+\frac{\beta_2 \chi '(r)}{r}+\frac{2 \phi '(r)}{r}+\frac{1}{2} \beta_2 \chi ''(r)+\phi ''(r) &=& 0 , \label{eq1_m1} \ \ \ \ \ \\
-m^2 \phi_A(r)+\frac{\rho_0 \phi_A(r)}{M_{pl}^2}-\frac{2 \beta_3 \phi_A(r) \chi '(r)}{r}+\frac{2 \phi_A'(r)}{r}-\beta_3 \phi_A(r) \chi ''(r)+\phi_A''(r) &=& 0 , \label{eq2_m1} \ \ \ \ \ \\
m^2 r^2 \chi '(r)+\frac{\beta_3 \rho_0 r^3 \phi_A(r)^2}{6 M_{pl}^2}+\frac{1}{2} \beta_2 r^2 \phi '(r)+\beta_3 r^2 \phi_A(r) \phi_A'(r)+2 \beta_3 r \chi '(r)^2 &=& 0 , \label{eq3_m1}\ \ \ \ \ 
\end{eqnarray}
and using equation \eqref{eq3_m1} with $m=0$
\begin{equation}
   \chi '(r) =\sqrt{- \dfrac{r}{12 \beta_3} \left(\dfrac{\beta _3 \rho _0 r \phi _A(r){}^2}{M_{\text{pl}}^2}+6 \beta _3 \phi _A(r) \phi _A'(r)+3 \beta _2 \phi '(r)\right)}.
\label{chiprime}
\end{equation}
We assume that $\phi(r)$ and $\phi_A(r)$ can be expressed as their background values plus a small perturbation
\begin{eqnarray}
\phi(r) &=& \phi_{0} + f_1(r), \ \ \ \ \ \ \text{with} \ \ \phi_{0} \gg f_1(r), \label{perturbed2}\\
\phi_A(r) &=& \phi_{A0} + f_2(r), \ \ \ \ \  \text{with} \ \ \phi_{A0} \gg f_2(r), \label{perturbed21}
\label{perturbed}
\end{eqnarray} 
where we only work with decreasing functions $\phi(r)$ and $\phi_A(r)$, or in other words, we assume $\phi'(r)<0$ and $\phi'_A(r)<0$. For a potential $V=0$ and $m=0$, equations \eqref{eq1_m1} and \eqref{eq2_m1} result in the following 
\begin{eqnarray}
 \frac{2 \alpha _4 \rho _0 \phi _0}{9 M_{\text{pl}}^2}r^3  + \dfrac{1}{2} \beta_2 r^2 \chi'(r) + r^2 f_1'(r) &=& C_1, \\
\frac{\rho_0 \phi_{A0}}{3 M_{pl}^2} r^3 - \phi_{A0} \beta_3 r^2 \chi'(r) + r^2 f_2'(r) &=& C_2,
\end{eqnarray}
and fixing $C_1=C_2=0$ we get
\begin{eqnarray}
 \frac{2 \alpha _4 \rho _0 \phi _0}{9 M_{\text{pl}}^2}r^3  + \dfrac{1}{2} \beta_2 r^2 \sqrt{- \dfrac{r}{12 \beta_3} \left(\dfrac{\beta _3 \rho _0 r \phi _{A0}^2}{M_{\text{pl}}^2}+6 \beta _3 \phi _{A0} f_2'(r)+3 \beta _2 f_1'(r)\right)} + r^2 f_1'(r) &=& 0, \nonumber\\
&& \\
\frac{\rho_0 \phi_{A0}}{3 M_{pl}^2} r^3 - \phi_{A0} \beta_3 r^2 \sqrt{- \dfrac{r}{12 \beta_3} \left(\dfrac{\beta _3 \rho _0 r \phi _{A0}^2}{M_{\text{pl}}^2}+6 \beta _3 \phi _{A0} f_2'(r)+3 \beta _2 f_1'(r)\right)} + r^2 f_2'(r) &=& 0, \nonumber\\
&&
\end{eqnarray}
Now, if we consider $f_1 =B_1 r^2$ and $f_2 =B_2 r^2$, the two last equations can be solved, and we obtain the following values for $B_1$ and $B_2$
\begin{eqnarray}
B_1 &=& \frac{\beta _2 \rho _0 }{12 \beta _3  M_{\text{pl}}^2} \bigg(-\frac{4 \alpha _4 \beta _3 \phi _0}{3 \beta _2} + \mathcal{F}(\xi) \bigg), \label{B1}\\
B_2 &=& -\frac{\rho _0 \phi _{\text{A0}}}{6  M_{\text{pl}}^2} \bigg( 1 + \mathcal{F}(\xi) \bigg),
\label{B2}
\end{eqnarray}
where,
\begin{equation}
    \mathcal{F}(\xi) = \xi-\sqrt{\xi (s_0+\xi)}, 
\end{equation}
with
\begin{equation}
    \xi = \frac{3  M_{\text{pl}}^2 \left(4 \beta _3^2 \phi _{\text{A0}}^2 - \beta _2^2\right)}{16 \rho _0 }\ \ \ \text{and} \ \ \ s_0 = \frac{4 \beta _3 \left(2 \alpha _4 \beta _2 \phi _0+3 \beta _3 \phi _{\text{A0}}^2\right)}{3 \left(4 \beta _3^2 \phi _{\text{A0}}^2 - \beta _2^2\right)},
    \label{eq_xi}
\end{equation}
In this work, we restrict to the case $\xi>0$  which implies   $\left(4 \beta _3^2 \phi _{\text{A0}}^2 - \beta _2^2\right)>0$. The reason for this is that we are interested in studying deviations from the case when the derivative self-interaction of the vector field is dominant over the other interactions \cite{DeFelice:2011th}.  We also can express  \eqref{chiprime} in terms of the obtained solutions for $\phi(r)$ and $\phi_A(r)$ (Eqs. \eqref{perturbed2}, \eqref{perturbed}, \eqref{B1} and \eqref{B2}).
\begin{eqnarray}
\chi'(r) &=& \frac{r}{6} \sqrt{\frac{\rho _0   \left(2 \alpha _4 \beta _2 \phi _0+3 \beta _3 \phi _{\text{A0}}^2\right)\left(1+\dfrac{2 \mathcal{F}(\xi)}{s_0}\right)}{\beta _3 M_{\text{pl}}^2}}. \label{chi_mr0}
\end{eqnarray} 
Considering the limit $\xi \ll 1$, the expression of the fields reduces to
\begin{eqnarray}
\phi (r) &\simeq&\phi_{0} -  \frac{ \alpha _4\phi _0\rho _0 }{9  M_{\text{pl}}^2}  r^2, \label{phi_xip}\\
\phi _A(r) &\simeq& \phi_{A0} -\frac{\rho _0 \phi _{\text{A0}}}{6  M_{\text{pl}}^2}  r^2, \label{phiA_xip}\\
\chi'(r) &\simeq& \frac{r}{6} \sqrt{\frac{\rho _0   \left(2 \alpha _4 \beta _2 \phi _0+3 \beta _3 \phi _{\text{A0}}^2\right)}{\beta _3 M_{\text{pl}}^2}}, \label{chi_xip}
\end{eqnarray} while for  $\xi \gg 1$, which is the case where the self-interaction of the vector field times the background value of its time component is greater than the interaction between the scalar and the vector field, we obtain
\begin{eqnarray}
\phi (r) &\simeq&\phi_{0} -  \frac{\beta _2 \rho _0 }{12 \beta _3  M_{\text{pl}}^2} \bigg(\frac{4 \alpha _4 \beta _3 \phi _0}{3 \beta _2} + \dfrac{s_0}{2} \bigg) r^2,\label{phi_xig}\\
\phi _A(r) &\simeq& \phi_{A0} -\frac{\rho _0 \phi _{\text{A0}}}{6  M_{\text{pl}}^2} \bigg( 1 - \dfrac{s_0}{2} \bigg) r^2, \label{phiA_xig}\\
\chi'(r) &\simeq& \frac{\rho _0  s_0}{6 \beta _3  M_{\text{pl}}^2} r. \label{chi_xig}
\end{eqnarray}
The following condition $\frac{s_0}{2}<1$ must be fulfilled for $\phi _A'(r)<0$. From these results, it is straightforward to deduce that the amplitude of $\chi'(r)$ in \eqref{chi_xip} is about $(\xi/s_{0})^{1/2}$ times smaller than the amplitude obtained in \eqref{chi_xig}. For $\beta_{2}=0$, that is $s_{0}=1$, we recover the result found in Ref. \cite{DeFelice:2016cri}, but it is crucial to notice that this reference does not include a scalar field. In this latter case, for $\left|\beta_{3}\right|\gg 1$, the screening mechanism works efficiently, and then the propagation of the longitudinal mode $\chi$ is suppressed. In the presence of scalar-vector interaction $\beta_2\neq 0$, this result remains correct as long as $s_{0}<2$.


\subsection{Solutions for \texorpdfstring{$r\ge r_*$}{Lg}}
Outside the body, using the same hypotheses as before ($V=0$ and $m=0$) and GR potentials \eqref{GR_potential_outside} as leading order, we obtain from equations \eqref{eomphi}, \eqref{eomA0} and \eqref{eomA1} the following expressions,
\begin{eqnarray}
 \dfrac{1}{2} \beta_2 r^2 \chi'(r) + r^2 \phi'(r) &=& - \frac{2 \alpha _4 \rho _0 \phi _0}{9 M_{\text{pl}}^2}r_*^3, \\
 - \phi_{A0} \beta_3 r^2 \chi'(r) + r^2 \phi_A'(r) &=& - \frac{\rho_0 \phi_{A0}}{3 M_{pl}^2} r_*^3. 
\end{eqnarray}
Defining 
\begin{equation}
    \mathcal{F}(s) = s-\sqrt{s (s_0+s)}, 
\end{equation}
with
\begin{equation}
    s = \xi\left(\frac{r}{r_*}\right)^3,
\end{equation}
we find
\begin{eqnarray}
\phi '(r) &=&\frac{\beta _2 \rho _0 r_*^3}{6 \beta _3 r^2 M_{\text{pl}}^2} \bigg(-\frac{4 \alpha _4 \beta _3 \phi _0}{3 \beta _2} + \mathcal{F}(s) \bigg),\label{phi_prime}\\
\phi _A'(r) &=& -\frac{\rho _0 r_*^3 \phi _{\text{A0}}}{3 r^2 M_{\text{pl}}^2} \bigg( 1 + \mathcal{F}(s) \bigg) ,\label{A0_prime}\\
\chi'(r) &=& \frac{1}{6} \sqrt{\frac{\rho _0 r_*^3  \left(2 \alpha _4 \beta _2 \phi _0+3 \beta _3 \phi _{\text{A0}}^2\right)\left(1+\dfrac{2 \mathcal{F}(s)}{s_0}\right)}{\beta _3 r M_{\text{pl}}^2}}.\label{chi_prime}
\end{eqnarray}
We observe that the behaviour of \eqref{phi_prime}, \eqref{A0_prime}, \eqref{chi_prime} changes when $s=1$. Therefore, we will identify the corresponding radius with $r_v$ such that $s=\left(r/r_{v}\right)^3$ \footnote{This behavior is similar to the one that appears in Galileon models. In these kind of models $r_v$ is named as the Vainshtein radius.}:

\bea
    && r_v = \left(\frac{16 \rho _0 r_*^3}{3  M_{\text{pl}}^2 \left(4 \beta _3^2 \phi _{\text{A0}}^2 - \beta _2^2\right)}\right)^{1/3},\nonumber\\
    &&r_v = \frac{r_{*}}{\xi^{1/3}}, \label{radio_v_xi} \label{rv}
\eea 

Now, we will show the behavior of the obtained solutions by taking limits on the value of $s$.  Taking the limit  $s \gg 1$, which implies that $r \gg r_v $  we obtain
\begin{eqnarray}
\phi '(r) &\simeq&-\frac{\beta _2 \rho _0 r_*^3}{6 \beta _3 r^2 M_{\text{pl}}^2} \bigg(\frac{4 \alpha _4 \beta _3 \phi _0}{3 \beta _2} +\frac{s_0}{2} \bigg),\label{phi_mar0}\\
\phi _A'(r) &\simeq& -\frac{\rho _0 r_*^3 \phi _{\text{A0}}}{3 r^2 M_{\text{pl}}^2} \bigg( 1 -\frac{s_0}{2} \bigg), \label{phiA_mar0}\\
\chi'(r) &\simeq& \frac{\rho _0 r_*^3 s_0}{6 \beta _3 r^2 M_{\text{pl}}^2}. \label{chi_mar0}
\end{eqnarray}

Otherwise, if we consider $s\ll 1$, which is equivalent to $r^* < r \ll r_v$ we get
\begin{eqnarray}\label{derivsmm1}
\phi '(r) &\simeq& -\frac{2 \alpha _4 \rho _0 r_*^3 \phi _0}{9 r^2 M_{\text{pl}}^2}, \label{phi_marv}\\
\phi _A'(r) &\simeq& -\frac{\rho _0 r_*^3 \phi _{\text{A0}}}{3 r^2 M_{\text{pl}}^2},  \label{phiA_marv}\\
\chi'(r) &\simeq& \frac{1}{6} \sqrt{\frac{\rho _0 r_*^3 \left(2 \alpha _4 \beta _2 \phi _0+3 \beta _3 \phi _{\text{A0}}^2\right)}{\beta _3 r M_{\text{pl}}^2}}.  \label{chi_marv}
\end{eqnarray}

 So, in the regime $r\gg r_{v}$, the longitudinal mode decreases faster than in the case $r_{*}<r\ll r_{v}$, and the condition $s_0<2$ ensures that the respective amplitudes are small. 
This particular behavior was also found in Ref. \cite{DeFelice:2016cri} for a vector-tensor theory. 

In this way, Eqs. \eqref{chi_xip}, \eqref{chi_xig}, \eqref{chi_mar0} and \eqref{chi_marv} show that the propagation of the longitudinal mode is suppressed inside and outside the compact body provided the condition  $s_{0}<2$ is fulfilled. This result extends the previous one found in Ref. \cite{DeFelice:2016cri} for generalized Proca theories to the case of SVT theories in the presence of scalar-vector interactions. 

Next, we will show that the requirements on  $\chi'(r)$ and  $\phi_A(r)$, imply bounds on $\alpha_4 \phi_0$ that depend on the couplings $\beta_2$ and $\beta_3$ and the vector background value $\phi_{A_0}$.
In fact, when $\xi\ll 1 $ and $s\ll 1$, it is necessary that $s_0\geq 0$ for $\chi'(r)$ to be a real number. Additionally, when $\xi\gg 1$ and $s\gg 1$, and for $\phi_A(r)$ to be a decreasing function, it is necessary that $s_0<2$. Therefore,  with $0\le s_0 < 2$ and taking $\beta_2$, $\beta_3$ and $\phi_{\text{A0}}$ as positive numbers, we obtain
\begin{equation}
 -\frac{3 \beta_3 \phi_{\text{A0}}^2}{2 \beta_2}\leq \alpha_4\phi_0\leq \frac{6 {\beta_3}^2 \phi_{\text{A0}}^2-3 {\beta_2}^2}{4 \beta_2\beta_3}.
 \label{conddes0}
\end{equation}
It follows from the last equation that if the condition  $\beta_2<\sqrt{2}\beta_3\phi_{\text{A0}}$ is fulfilled, the upper bound of $\alpha_4\phi_0$ is positive, while when it is not, $\alpha_4\phi_0$ is bounded between two negative values.


\section{Numerical solutions}\label{Num_Res}

In this section, we check that the approximate analytical solutions obtained in the previous section are continuous at $r=r*$. One of the aims of this work is to test our model with solar systems constraints.  Therefore, we focus on the case where the source body is the Sun and numerically solve equations \eqref{eom00}-\eqref{eomA1}.   We consider a more realistic model for the solar density, taking $\rho_S (r) = \rho_0 e^{-a r^2/r^2_S}$. Here $a$ is of order 1, $r_S$ refers the Sun radius and $\rho_0=162.2$ $\rm{g/m^{3}}$ represents the solar central density.


Also, for numerical purposes, we introduce the variables
\begin{equation}
    x = \dfrac{r}{r_*}, \ \ \ \ \ \ \ \ \ y_0 = \dfrac{\phi}{\phi_0}, \ \ \ \ \ \ \ \ \ y = \dfrac{\phi_A}{\phi_0},  \ \ \ \ \ \ \ \ \ z = \dfrac{\chi'}{\phi_0} , \label{numerical_definitions}
\end{equation}
and we consider $\phi_0=\phi_{A0}= \chi'_0$ at $r=0$. Thus, expressions \eqref{eomphi}, \eqref{eomA0} and \eqref{eomA1} evaluated at the particular model described by \eqref{particular_model} and fixing $V=m^2=0$, result in the following equations:
\begin{eqnarray}
    x \left(\beta _2 r_* \left(x z'+z \left(-x \Phi '+x \Psi '+2\right)\right)+y_0' \left(-2 x \Phi '+2 x \Psi '+4\right)+2 x y_0''\right) && \nonumber\\
    -8 \alpha _4 y_0 \left(-e^{2 \Phi }+x^2 \Psi ''+x^2 \left(\Psi '\right)^2-x \Phi ' \left(x \Psi '+2\right)+2 x \Psi '+1\right) &=& 0, \\
    -\frac{2 \beta _3 r_* y z \phi _0}{x}-\beta _3 r_* y \phi _0 z'+\beta _3 r_* y z \phi _0 \Phi '-\beta _3 r_* y z \phi _0 \Psi '+\frac{2 y'}{x}+\frac{4 y \Psi '}{x}+y''&&\nonumber\\
    -\Phi ' y'+3 \Psi ' y'-2 y \Phi ' \Psi '+2 y \Psi ''+2 y \left(\Psi '\right)^2 &=&0, \\
    x \Psi ' \left(y^2 e^{2 (\Phi +\Psi )}+z^2\right)+x y e^{2 (\Phi +\Psi )} y'+\frac{\beta _2 e^{2 \Phi } x y_0'}{2 \beta _3 \phi _0}+2 z^2 &=& 0.
\end{eqnarray}

\begin{figure}[ht]%
    \centering
    \subfloat[\centering $\alpha_4=10^{-6}$, $\xi=1$ and $\frac{\beta_2}{\beta_3 \phi_0} = 1$]{{\includegraphics[width=7.5cm]{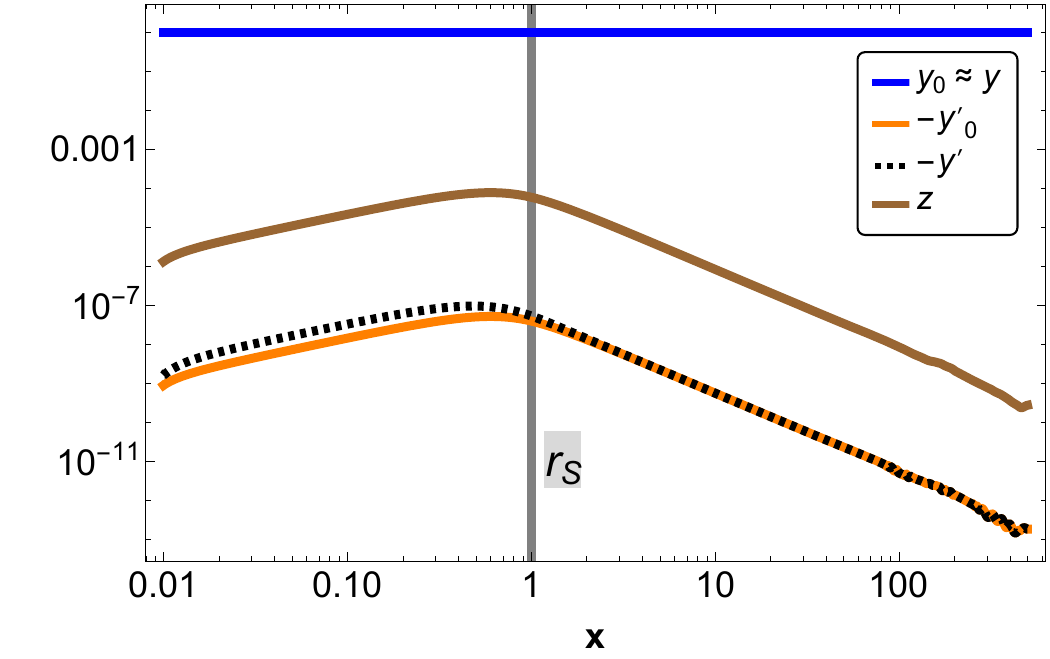} }}%
    \subfloat[\centering $\alpha_4=10^{-6}$, $\xi=10^{-4}$ and $\frac{\beta_2}{\beta_3 \phi_0} = 1$]{{\includegraphics[width=7.5cm]{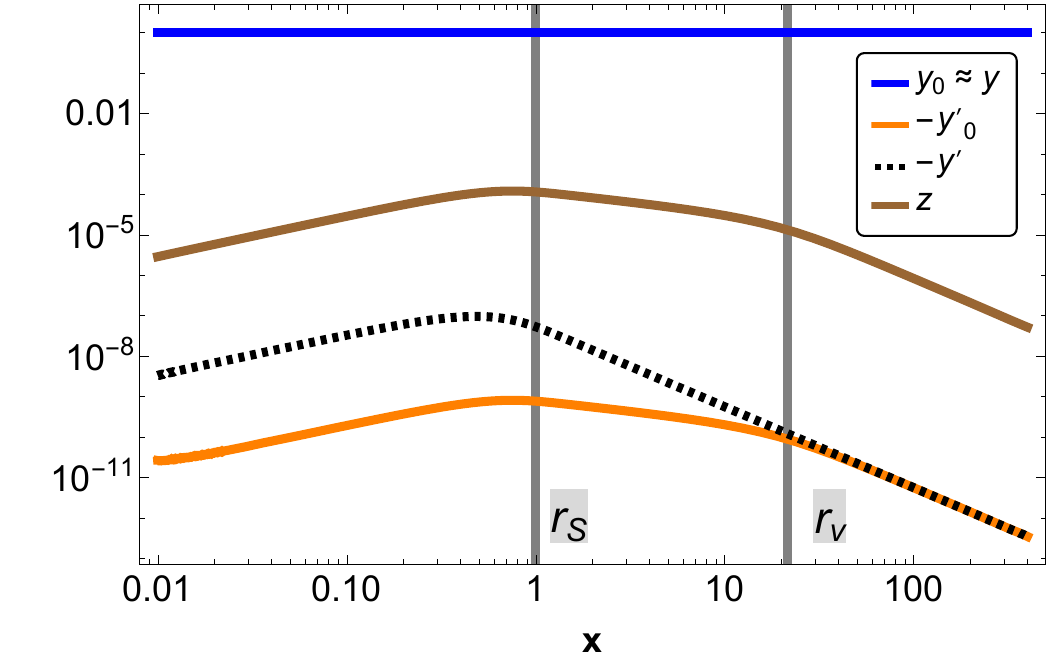} }}%
    \qquad
    \subfloat[\centering $\alpha_4=1 $, $\xi=1$ and $\frac{\beta_2}{\beta_3 \phi_0} = 1$]{{\includegraphics[width=7.5cm]{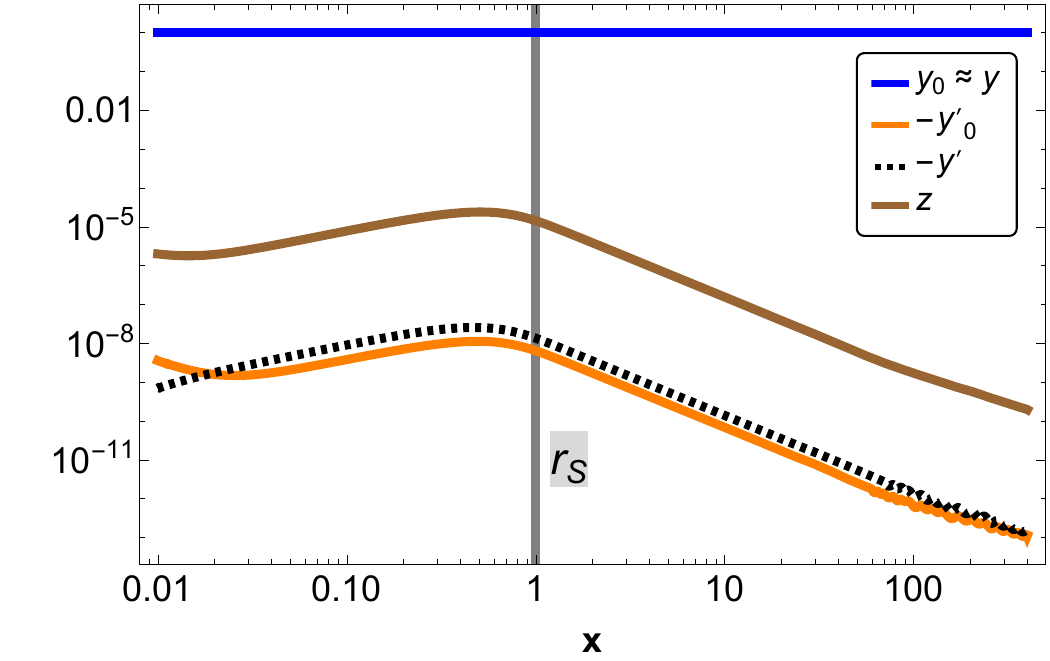} }}%
    \subfloat[\centering $\alpha_4=1$, $\xi=10^{-4}$ and $\frac{\beta_2}{\beta_3 \phi_0} = 1$]{{\includegraphics[width=7.5cm]{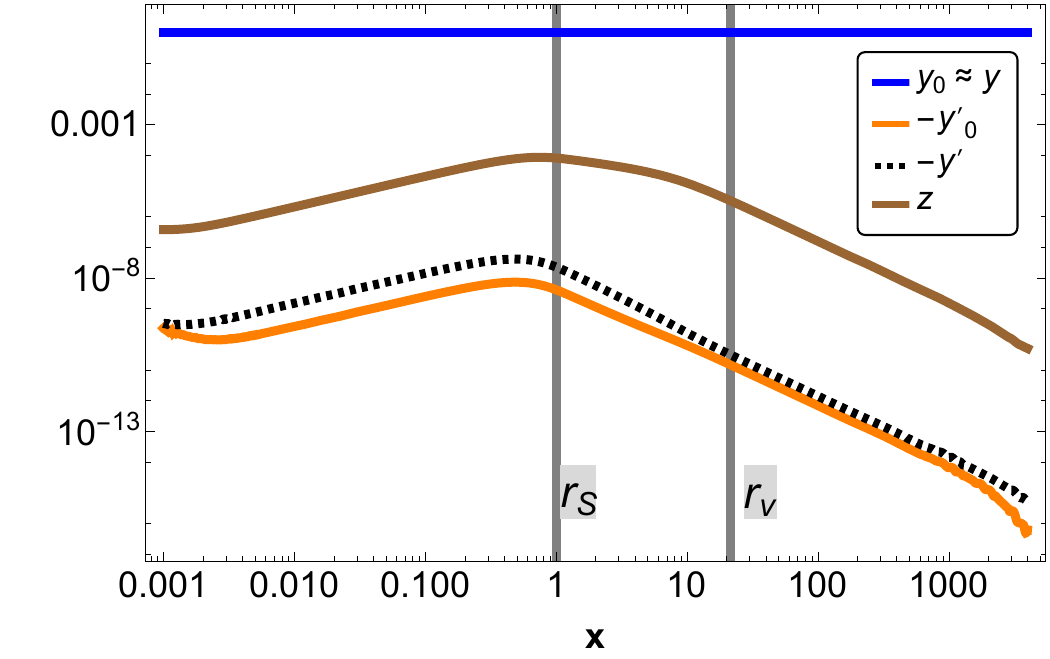} }}%
    \qquad
    \subfloat[\centering $\alpha_4=10^{-6}$, $\xi=1$ and $\frac{\beta_2}{\beta_3 \phi_0} = 10^{-1}$]{{\includegraphics[width=7.5cm]{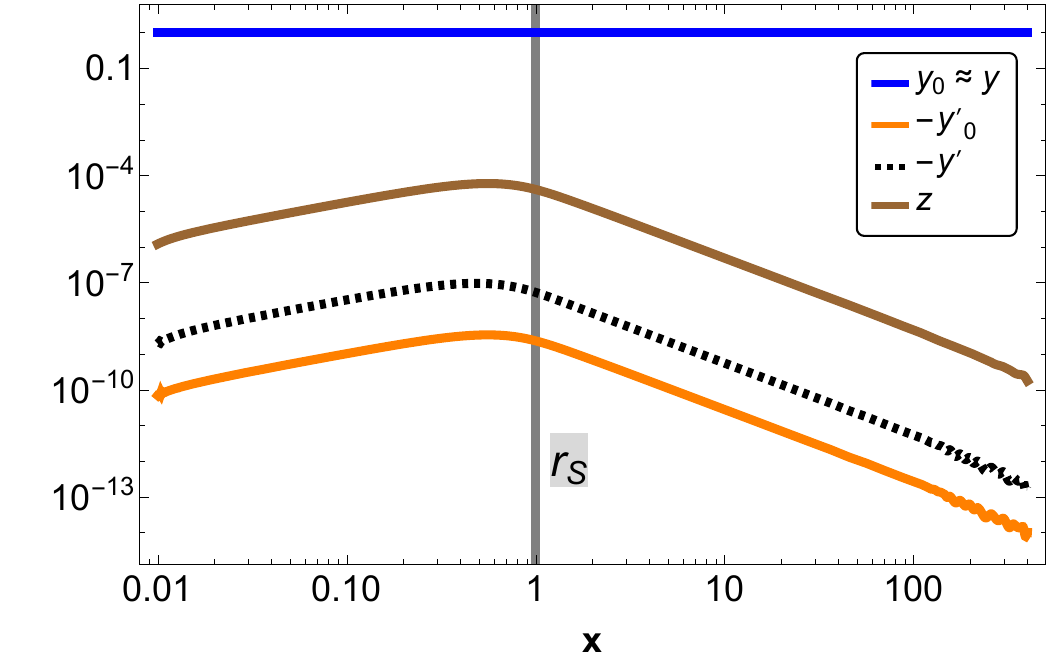} }}%
    \subfloat[\centering $\alpha_4=10^{-6}$, $\xi=10^{-4}$ and $\frac{\beta_2}{\beta_3 \phi_0} = 10^{-1}$]{{\includegraphics[width=7.5cm]{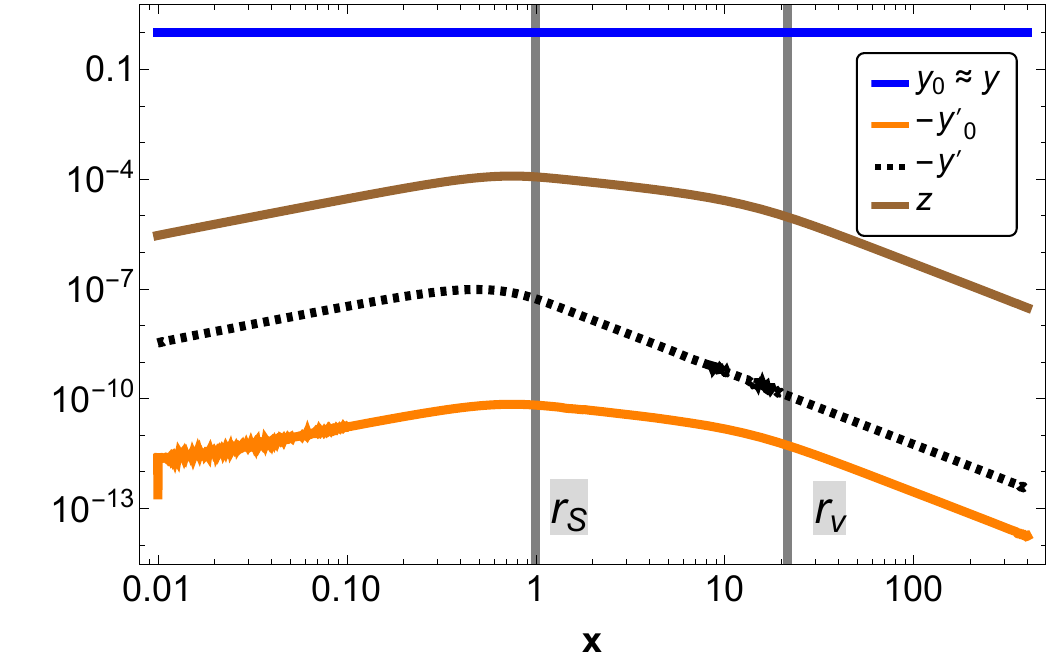} }}%
    \qquad
    \caption{\scriptsize Numerical solutions of $y_0 (x)$, $y_0'(x)$, $y(x)$, $y'(x)$ and $z(x)$  as functions of $x=\dfrac{r}{r_S}$ for different values of $\alpha_4$, $\xi$ and $\dfrac{\beta_2}{\beta_3 \phi_0}$. We consider $a=4$ and $\Phi_0 = 10^{-6}$. Vertical lines represent $r=r_S$ and $r=r_V$. Note that the cases considered in the left column, $r=r_S=r_V$.}%
    \label{fig:numerical}%
\end{figure}

In Fig \ref{fig:numerical} we depict the behavior of $y_0$, $y_0'$, $y$, $y'$ and $z$  for several different values of the parameters $\alpha_{4}$, $\xi$ and the quotient  $\beta_{2}/(\beta_{3} \phi_{0})$, considering fixed values of $\Phi_{0}=10^{-6}$ (which is the approximate value of the gravitational potential of the Sun at its surface) and $a=4$. In order to determine the boundary conditions of $y_{0}$, $y$, and $dy/dx$ around the center of the body (in this case, the Sun), we use Eqs. \eqref{perturbed2}, \eqref{perturbed},  \eqref{chi_mr0}. It follows from  Fig. \ref{fig:numerical}  that the functions $-\phi'$, $-\phi_{A}'$ and $\chi'$  grow linearly in $r$ for the distance smaller than $r_S$ as can be expected from the analytical solutions obtained in the limit $r<r_*$ (\eqref{phi_xip}-\eqref{chi_xip}). Likewise, for $r>r_S$ Fig. \ref{fig:numerical}  shows a decreasing behavior for the same functions, which is also consistent with the obtained analytical solutions (Eqs.\eqref{phi_mar0}-\eqref{chi_marv}). Also, from \eqref{radio_v_xi}, we can see a relation between  $r_v$, parameter $\xi$ and $r_S$; in particular, for $\xi=1$, we obtain $r_S=r_V$. Fig. \ref{fig:numerical}   shows that the functions $\phi(r)$ and $\phi_{A}(r)$ are nearly constants in the whole regime studied. The reason for this lies in that the $r$-dependent corrections to $\phi(r)$ and $\phi_{A}$ are at most of  order $\phi_0 \Phi_0$ and $\phi_{A0} \Phi_0$ .

\section{Corrections to gravitational potentials}\label{subsecII}

In this section, we calculate the corrections to the gravitational potentials of GR,  $\Phi_{GR}$ and $\Psi_{GR}$, that arise when considering an SVT model described by \eqref{particular_model}. We  focus in the gravitational potential of the Sun ($\Phi_0=\dfrac{\rho_0 r_S^2}{M_{\text{pl}}^2}\sim 10^{-6}$). The first-order corrections to the gravitational field equations \eqref{eom00} and \eqref{eom11} give us

\begin{eqnarray}
\dfrac{2 M_{\text{pl}}^2}{r} \Phi' - \dfrac{M_{\text{pl}}^2}{r^2}\left( 1- e^{2 \Phi} \right) &=&  e^{2 \Phi}\rho_m + \Delta_{\Phi} , \label{eq_corrections1}\\
\dfrac{2 M_{\text{pl}}^2}{r} \Psi' + \dfrac{M_{\text{pl}}^2}{r^2}\left( 1- e^{2 \Phi} \right) &=&  e^{2 \Phi}P_m + \Delta_{\Psi},  \label{eq_corrections2}
\end{eqnarray} where the functions $\Delta_{\Phi}$ and $\Delta_{\Psi}$ represent the first order corrections. 
We stress that the regime $\xi \ll 1$ is very unlikely for a non-vanishing vector field because according to Eq.\eqref{eq_xi} the difference $\beta_3^ 2 \phi_{A0}^ 2 - \beta_2^2$ must be of order $10^{-94} M_{\text{pl}}^2$ if $\rho_0$ is the density of the Sun. This would require an extremely fine-tuning of the model parameters. Therefore, we focus on the regime $\xi \gg 1$. First we obtain   the expressions of $\Delta_{\Phi}$ and $\Delta_{\Psi}$  inside the sun ($r<r_S$). In this  regime, we can approximate the solutions for $\phi'(r)$, $\phi'_A(r)$, and $\chi'(r)$ with equations \eqref{phi_xig}, \eqref{phiA_xig}, and \eqref{chi_xig}. Consequently, when considering the leading-order potentials as given in \eqref{GR_potential_outside}, the first-order corrections for \eqref{eq_corrections1} and \eqref{eq_corrections2} can be expressed as follows
\begin{equation}
    \Delta_{\Phi} \simeq \frac{\rho _0 \left(4 \alpha _4 \left(4 \alpha _4+3\right) \beta _3 \phi _0^2+s_0 \left(6 \alpha _4 \beta _2 \phi _0-3 \beta _3 \phi _{\text{A0}}^2\right)\right)}{6 \beta _3 M_{\text{pl}}^2}, \label{corrections21}
\end{equation}
and
\begin{equation}
    \Delta_{\Psi} \simeq -\frac{2 \alpha _4 \phi _0 \rho _0 \left(8 \alpha _4 \beta _3 \phi _0+3 \beta _2 s_0\right)}{9 \beta _3 M_{\text{pl}}^2} . \label{corrections22}
\end{equation}

In what follows, we will show the conditions that make these corrections of order $10^{-2} \rho_0$ so that the potentials inside the Sun can be considered to be the same as in GR.

Now, we focus on the case $r>r_s$. For this, we use leading-order potentials $\Phi_{GR}$ and $\Psi_{GR}$, Eq. \eqref{GR_potential_outside}, and solutions of $\phi'(r)$, $\phi'_A(r)$ and $\chi'(r)$ given by \eqref{phi_prime}, \eqref{A0_prime} and \eqref{chi_prime}. To analyze this regime, it is useful to divide the analysis in two cases of interest: $s = \xi\left(\dfrac{r}{r_S}\right)^3\gg 1$ and $s = \xi\left(\dfrac{r}{r_S}\right)^3\ll 1$.

\subsection{The \texorpdfstring{$s\gg 1$}{Lg} regime}\label{smuchomayorauno}

If we consider the case $\xi \ge 1$, 
the following relation
\begin{equation}
\label{relation}
4\beta_3^2\phi_{A0}^2-\beta_2^2\gtrsim 5\times 10^{-6}r_S^{-2},
\end{equation}
is satisfied, and it follows from \eqref{rv} that  $r_v\leq r_S$.

In this regime, and assuming that $s \gg 1$ (which implies $r\gg r_v$), the solutions of $\phi'(r)$, $\phi'_A(r)$ and $\chi'(r)$ are approximately given by \eqref{phi_mar0}, \eqref{phiA_mar0} and \eqref{chi_mar0}. Consequently, considering leading-order potentials \eqref{GR_potential_outside}, the first order corrections of equations \eqref{eq_corrections1} and \eqref{eq_corrections2}  can be written as

\begin{equation}
    \Delta_{\Phi} \simeq \frac{r_S^2 \Phi _0^2 \phi _{\text{A0}}^2 s_1}{72  r^4} \ \ \text{and} \ \ \Delta_{\Psi} \simeq -\frac{r_S \Phi _0 \phi _{\text{A0}}^2 s_2}{ r^3} - \frac{r_S^2 \Phi _0^2 \phi _{\text{A0}}^2 s_3}{72 r^4}, \label{corrections1}
\end{equation}
where,
\begin{eqnarray}
s_1 &=&  -\dfrac{1}{\beta _2^2}\left(\beta _2^2 \left(2-7 s_0\right) s_0-4 \left(8 \alpha _4-5\right) \beta _3^2 \left(s_0-1\right){}^2 \phi _{\text{A0}}^2\right), \\
s_2 &=&  \dfrac{1}{\beta _2^2}\left(s_0-1\right) \left(\beta _2^2 s_0-4 \beta _3^2 \left(s_0-1\right) \phi _{\text{A0}}^2\right), \\
s_3 &=&  \dfrac{1}{\beta _2^2}\left(\beta _2^2 s_0 \left(7 s_0-6\right)-28 \beta _3^2 \left(s_0-1\right){}^2 \phi _{\text{A0}}^2\right).
\end{eqnarray}
After integrating \eqref{eq_corrections1} and \eqref{eq_corrections2}, along with equation \eqref{corrections1}, we find that outside the Sun, the gravitational potentials are given by
\begin{eqnarray}
\Phi(r) &\simeq& \frac{r_S \Phi _0}{6 r} \left[1-\frac{r_S  s_1 \Phi _0 \phi _{\text{A0}}^2 }{24 M_{\text{pl} }^2 r} \right], \\
\Psi(r) &\simeq& -\frac{r_S \Phi _0}{6 r} \left[1-\frac{3 s_2 \phi _{\text{A0}}^2}{M_{\text{pl}}^2} -\frac{r_S \left(s_1+s_3\right) \Phi _0 \phi _{\text{A0}}^2}{48 M_{\text{pl}}^2 r} \right].
\end{eqnarray}
In addition,  the PPN parameter $\gamma \equiv - \Phi/\Psi$ becomes
\begin{equation}
\label{gamma11}
    \gamma_{\text{th}} \simeq 1 - \frac{3 s_2 \phi _{\text{A0}}^2}{M_{\text{pl}}^2} + \frac{r_S \Phi _0 \phi _{\text{A0}}^2}{48 M_{\text{pl}}^2 r} \left( -\frac{6 s_1 s_2 \phi _{\text{A0}}^2}{M_{\text{pl}}^2}+s_1-s_3 \right).
\end{equation}

\begin{figure}[ht]%
    \centering
     \subfloat[\centering]
    {{\includegraphics[width=8cm]{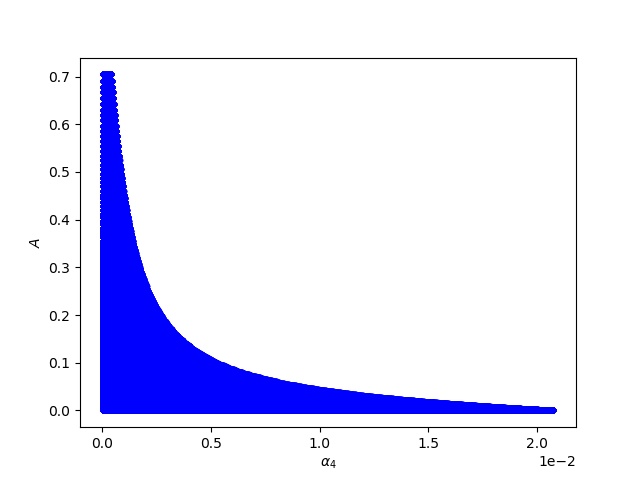} }}%
    \subfloat[\centering]
    {{\includegraphics[width=8cm]{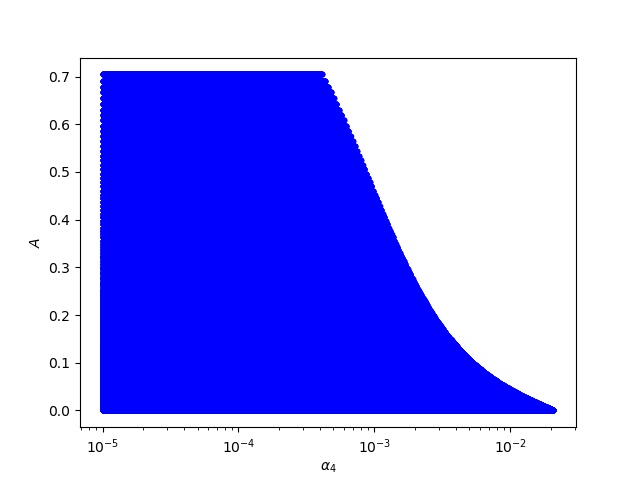} }}%
    \qquad
    \caption{The parameter region that satisfies the observational bound for $|\gamma-1|$ when $0<\alpha_4<\frac{9-16A^2}{12A}$ and $0\le A<0.7$.  In order to better understand the behaviour,  subfigure (a) has a linear scale in $\alpha_4$ while subfigure (b), has a logaritmic scale in $\alpha_4$. }%
    \label{fig:numericalalphapos}%
\end{figure}

 Since $\xi>0$, it follows that $4\beta_3^2\phi_{A0}^2-\beta_2^2>0$  . Therefore, we can define $A=\frac{\beta_2}{2\beta_3\phi_{A0}}$ so that  $0<A<1$ and rewrite \eqref{gamma11} in terms of this new parameter:
\begin{align}
\label{gamma_total}
\begin{split}
\Big|\gamma-1\Big|_{\text{th}}  &\simeq -\frac{4 \text{$\alpha_4 $} \text{$\phi_0$} (3 A \phi_{\text{A0}}+4 \text{$\alpha_4$} \text{$\phi_0$})}{3 \left(A^2-1\right) M_{\text{pl}}^2}
+\frac{r_S \Phi _0 }{48  r}\Big[
\frac{32 \text{$\alpha_4$}^2 (4 \text{$\alpha_4$}+1)  \text{$\phi_0$}^2}{9 \left(A^2-1\right)^2 M_{\text{pl}}^2}-\frac{512 \text{$\alpha_4$}^4  \text{$\phi_0$}^4 \left(7 A^2+8 \text{$\alpha_4$}-5\right)}{27 \left(A^2-1\right)^3 M_{\text{pl}}^4}\\ &- \frac{16 A \text{$\alpha_4$}  \text{$\phi_0$} \text{$\phi_{\text{A0}} $} \left(A^4-4 A^2 \text{$\alpha_4$}+4 \text{$\alpha_4$}+2\right)}{3 \left(A^2-1\right)^3 M_{\text{pl}}^2}
+\frac{2  \text{$\phi_{\text{A0}}$} \left(8 A^3 \text{$\alpha_4$} \text{$\phi_0$}+\left(A^2-1\right) \text{$\phi_{\text{A0}} $} \left(A^2 (4 \text{$\alpha_4$}-1)+2\right)\right)}{\left(A^2-1\right)^3 M_{\text{pl}}^2}\\
&-\frac{32 \text{$\alpha_4$}^2  \text{$\phi_0$}^2 \text{$\phi_{\text{A0}} $} \left(\left(2 A^4-A^2+5\right) \text{$\phi_{\text{A0}} $}+4 A \text{$\alpha_4$} (8 \text{$\alpha_4$}-1) \text{$\phi_0$}\right)}{3 \left(A^2-1\right)^3 M_{\text{pl}}^4}
-\frac{8 A \text{$\alpha_4$}  \text{$\phi_0$} \text{$\phi_{\text{A0}} $}^2 \left(32 A \text{$\alpha_4$}^2 \text{$\phi_0$}+5 \text{$\phi_{\text{A0}} $}\right)}{\left(A^2-1\right)^3 M_{\text{pl}}^4}\\
&-\frac{8 A^3 \text{$\alpha_4$}  \text{$\phi_0$} \text{$\phi_{\text{A0}} $} \left(16 \text{$\alpha_4$}^2 \text{$\phi_0$}^2+(8 \text{$\alpha_4$}-3) \text{$\phi_{\text{A0}} $}^2\right)}{\left(A^2-1\right)^3 M_{\text{pl}}^4}
\Big],
\end{split}    
\end{align} 
We want to compare the above expression with the experimental bound $|\gamma-1|\leq 2.3 \times 10^{-5}$ reported in \cite{2003Natur.425..374B}. First, we note that the Cassini mission measures the delay time of a signal traveling from Earth to Saturn  and back. Therefore, the relevant values of $r$ in \eqref{gamma_total} are in the range  $r_S< r < d_{\rm SAT} \simeq 10^3 \, r_S$ where $d_{\rm SAT} $ is the distance from the Sun to Saturn. Besides, since the theoretical expression described in  \eqref{gamma_total} is complex, we need to make some assumptions. Therefore, we consider $\beta_2$, $\beta_3$ $\phi_0$ and $\phi_{\text{A0}}$  all positive.  Next, we use a numerical method to obtain the regions in the  $A-\alpha_4$ plane that satisfy the following conditions: i) expression \eqref{gamma_total} is below the observational constraint and ii) the corrections described by Eqs.\eqref{corrections21} and \eqref{corrections22} are lower than $10^{-2} \rho_0$. We analyze two cases: i) $0\leq \alpha_4\phi_0\leq \frac{6 {\beta_3}^2 \phi_{\text{A0}}^2-3 {\beta_2}^2}{4 \beta_2\beta_3}$ and ii) $ -\frac{3 \beta_3 \phi_{\text{A0}}^2}{2 \beta_2}\leq \alpha_4\phi_0<0$. As regards the values of  $\phi_0$ and $\phi_{A0}$, a first reasonable choice is that both quantities are equal than Planck's mass. However, we found that no value of $\alpha_4$ and $A$ was able to satisfy condition ii) i.e. that  the corrections to the potentials are negligible inside the Sun. However, this last condition can be achieved if we take  $\phi_0\leq 0.1 M_{\text{pl}}$ and $\phi_{\text{A0}}\leq 0.1 M_{\text{pl}}$, so we have decided to fix both magnitudes in theses values ($\phi _0=\phi_{\text{A0}}= 0.1 M_{\text{pl}}$).
Fig. \ref{fig:numericalalphapos} shows the region in the $\alpha_4 -A$ plane that meets conditions i) and ii) in the case $0<\alpha_4<\frac{9-16A^2}{12A}$. We note that  $\alpha_4>0$,  necessarily implies  $A<\frac{\sqrt{2}}{2}$. Besides, for this case, $|\gamma -1|$ decreases while $r$ increases, so its maximum value is reached when $r=r_S$ and therefore the reported values in Fig. \ref{fig:numericalalphapos} are calculated taking $r=r_S$ in Eq. \eqref{gamma_total}.
 On the other hand, it follows from Fig. \ref{fig:numericalalphapos} that if  $|\alpha_4|\lesssim 5\times 10^{-4}$, all $A$ values in the proposed range fulfill the experimental condition for $|\gamma-1|$. While as $\alpha_4$ grows, the allowed region narrows very quickly, and the $A$ values get smaller (e.g for $\alpha_4\sim 5\times 10^{-5}$, $A<0.1$) in such a way that for the limit of A tending to 0, $\alpha_4<2.2\times 10^{-2}$.

\begin{figure}[ht]%
    \centering
    \subfloat[\centering]{{\includegraphics[width=8cm]{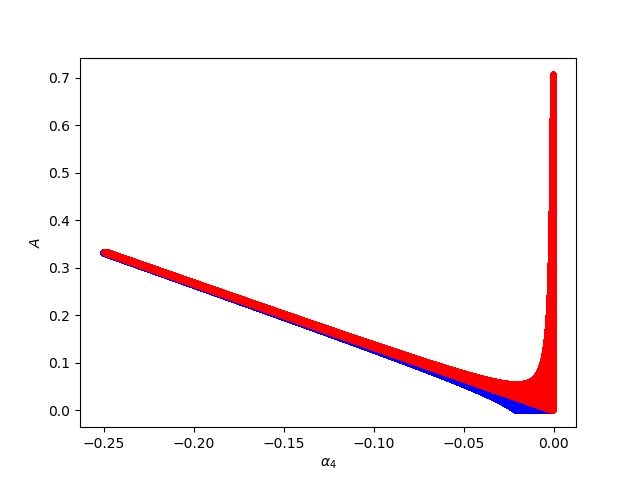} }}%
    \subfloat[\centering]{{\includegraphics[width=8cm]{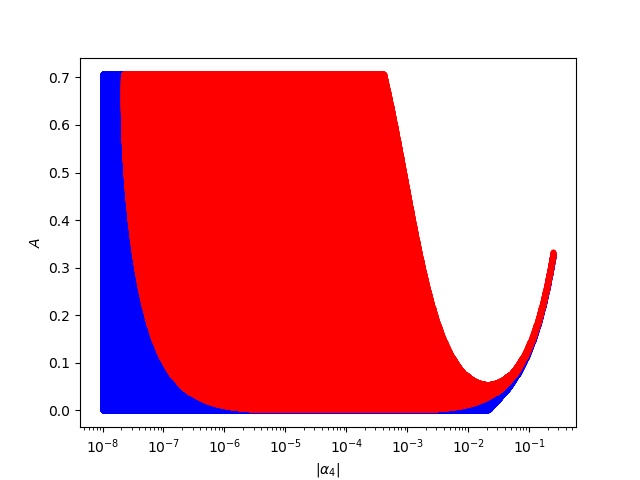} }}%
    \qquad
    \caption{The parameter region that satisfies the observational bound for $|\gamma-1|$ when  $-\frac{3}{4A}<\alpha_4<0$ and $0\le A<1$. The subfigure (b) represents the logarithmic mode of subfigure (a) in order to better understand the behavior of the variables when $\alpha_4$ approaches to 0. The blue regions symbolize the $\alpha_4$ and $A$ values that are consistent with the observational bound for $r=r_S$ while the red ones, for $r=d_{\text{SAT}}$.  }%
    \label{fig:numericalalpha}%
\end{figure}

Fig. \ref{fig:numericalalpha} shows the allowed region in the $\alpha_4 - A$ plane in the case $-\frac{3}{4A}<\alpha_4<0$. We note that if $\alpha_4 < 0$, $0<A<1$.
For certain values of $\alpha_4$ and $A$, the PPN parameter $\gamma$ is maximum when $r=r_S$ (blue region) while for others, it is reached when $r=d_{\text{SAT}}$ (red region). Besides, it follows from Fig. \ref{fig:numericalalpha} that if $|\alpha_4|\gtrsim 0.02$, the relationship between the variables is almost linear given by the approximate expression $\alpha_4\sim -A + 3 \times 10^{-2}$, being $\alpha_4\sim -0.25$ the minimum value for said variable. On the other hand,  if $|\alpha_4|\lesssim 0.02$ the behavior changes. Indeed, as $\alpha_4$ approaches 0 a greater number of $A$ values are possible to satisfy the constraint. In this way, if $|\alpha_4|\lesssim 1\times 10^{-3}$ all $A<0.7$ values in the proposed range fulfill the experimental condition for $|\gamma-1|$. We note that although the allowed range for $A$ is between $0$ and $1$, the observational constraints restrict the allowed values to $0<A<0.7$. 

\section{Concluding Remarks}\label{conclusion_f}

In the present paper, we have analysed the screening mechanism of the fifth force in scalar-vector-tensor (SVT) theories. These latter theories arise from unifying scalar Horndeski and generalized Proca theories and keeping the field equations at second order. For the broken $U(1)$ gauge symmetry, there is a propagating longitudinal scalar mode in addition to the other standard five degrees of freedom. Thus, to avoid the propagation of the fifth force and comply with local gravity constraints, a screening mechanism in SVT theories is required. The screening mechanism studied here is similar to the Vainshtein mechanism but based on both the derivative self-interactions of the vector field and the interactions of the scalar field with the vector field and curvature \cite{DeFelice:2016cri}.

In order to obtain analytical solutions for the fields, we carried out an analysis similar to the one presented in \cite{DeFelice:2016cri}. We assumed that the derivative self-interaction of the vector field dominates on the interaction between the fields.  
In this way, we studied a field configuration where the interaction between the scalar and vector fields is a small perturbation on the dynamics of this two-field system. This is a natural choice because our proposal relied on studying the field space dynamics around the model proposed in Ref. \cite{DeFelice:2016cri}. Thus, we found analytical solutions for the scalar and vector profiles by studying a gravitating spherically symmetric compact body. These solutions were obtained by first studying the region inside the body and then outside it. We found that for a dominant derivative self-interaction of the vector field, the screening mechanism works efficiently, and then the propagation of the additional longitudinal mode is suppressed. This particular behavior was also found in Ref. \cite{DeFelice:2016cri} for a vector-tensor theory. Furthermore, in the presence of scalar-vector interaction, this result remains correct as long as some conditions on the parameters are satisfied. So, we have shown that the propagation of the longitudinal mode is suppressed inside and outside the compact body. This result extends the previous one found in Ref. \cite{DeFelice:2016cri} for generalized Proca theories to the case of SVT theories in the presence of scalar-vector interactions. Finally, we corroborated all these analytical results by numerically integrating the field equations. 

Additionally, we have computed the corrections to the Post-Newtonian parameter $\gamma$. By applying the solar system constraints, we have set bounds for the
leading-order values of the vector field  
and the scalar field 
as well as for the non-minimal scalar-tensor coupling parameter. 
Our results are compatible with those presented by A. De Felice et al. \cite{DeFelice:2016cri} in the limit of a dominant derivative self-interaction of the vector field. 
Besides, we point out that the inclusion of the non-minimally coupled scalar field leads to additional deviations from GR in the PPN parameter $\gamma$.



Thus, we conclude that the present model, including both interactions, can avoid the propagation of the additional longitudinal mode arising in these theories. Therefore, it is also compatible with local gravity constraints. This result extends the previous one found in the literature for generalized Proca theories to the case of SVT theories in the presence of scalar-vector interactions.

\begin{acknowledgments}
M. Gonzalez-Espinoza acknowledges the financial support of FONDECYT de Postdoctorado, N° 3230801. G. Otalora acknowledges the financial
support of FONDECYT Grant 1220065.
 L. Kraiselburd and S. Landau are supported by CONICET grant PIP 11220200100729CO, grants  G175 from UNLP, and grant 20020170100129BA UBACYT.
\end{acknowledgments}


\bibliographystyle{spphys}       
\bibliography{bio}   

\begin{thebibliography}{10}
\providecommand{\url}[1]{{#1}}
\providecommand{\urlprefix}{URL }
\expandafter\ifx\csname urlstyle\endcsname\relax
  \providecommand{\doi}[1]{DOI \discretionary{}{}{}#1}\else
  \providecommand{\doi}{DOI \discretionary{}{}{}\begingroup \urlstyle{rm}\Url}\fi

\bibitem{Riess:1998cb}
A.G. Riess, et~al., Astron. J. \textbf{116}, 1009 (1998)

\bibitem{Perlmutter:1998np}
S.~Perlmutter, et~al., Astrophys. J. \textbf{517}, 565 (1999)

\bibitem{Zwicky:1933gu}
F.~Zwicky, Helv. Phys. Acta \textbf{6}, 110 (1933)

\bibitem{Weinberg:1988cp}
S.~Weinberg, Rev. Mod. Phys. \textbf{61}, 1 (1989)

\bibitem{Carroll:2000fy}
S.M. Carroll, Living Rev. Rel. \textbf{4}, 1 (2001)

\bibitem{Padilla:2015aaa}
A.~Padilla, arXiv 1502.05296 [hep-th]  (2015)

\bibitem{Abdalla:2022yfr}
E.~Abdalla, et~al., JHEAp \textbf{34}, 49 (2022)

\bibitem{DiValentino:2020zio}
E.~Di~Valentino, et~al., Astropart. Phys. \textbf{131}, 102605 (2021)

\bibitem{DiValentino:2020vvd}
E.~Di~Valentino, et~al., Astropart. Phys. \textbf{131}, 102604 (2021)

\bibitem{Heisenberg:2022lob}
L.~Heisenberg, H.~Villarrubia-Rojo, J.~Zosso, arXiv 2201.11623 astro-ph.CO  (2022)

\bibitem{Planckcosmo2018}
{Planck Collaboration}, {N. Aghanim}, {Y. Akrami}, {M. Ashdown}, {J. Aumont}, {C. Baccigalupi}, {M. Ballardini}, {A. J. Banday}, {R. B. Barreiro}, {N. Bartolo}, {S. Basak}, et~al., A\&A \textbf{641}, A6 (2020).
\newblock \doi{10.1051/0004-6361/201833910}.
\newblock \urlprefix\url{https://doi.org/10.1051/0004-6361/201833910}

\bibitem{Riess2022}
A.G. Riess, W.~Yuan, L.M. Macri, D.~Scolnic, D.~Brout, S.~Casertano, D.O. Jones, Y.~Murakami, G.S. Anand, L.~Breuval, et~al., The Astrophysical Journal Letters \textbf{934}(1), L7 (2022).
\newblock \doi{10.3847/2041-8213/ac5c5b}.
\newblock \urlprefix\url{https://dx.doi.org/10.3847/2041-8213/ac5c5b}

\bibitem{Heymans:2020gsg}
C.~Heymans, et~al., Astron. Astrophys. \textbf{646}, A140 (2021)

\bibitem{Nunes:2021ipq}
R.C. Nunes, S.~Vagnozzi, Mon. Not. Roy. Astron. Soc. \textbf{505}(4), 5427 (2021)

\bibitem{Heisenberg:2022gqk}
L.~Heisenberg, H.~Villarrubia-Rojo, J.~Zosso, Phys. Rev. D \textbf{106}(4), 043503 (2022)

\bibitem{Wetterich:1987fm}
C.~Wetterich, Nucl. Phys. B \textbf{302}, 668 (1988)

\bibitem{Ratra:1987rm}
B.~Ratra, P.~Peebles, Phys. Rev. D \textbf{37}, 3406 (1988)

\bibitem{Carroll:1998zi}
S.M. Carroll, Phys. Rev. Lett. \textbf{81}, 3067 (1998)

\bibitem{Tsujikawa:2013fta}
S.~Tsujikawa, Class. Quant. Grav. \textbf{30}, 214003 (2013)

\bibitem{Chiba:1999ka}
T.~Chiba, T.~Okabe, M.~Yamaguchi, Phys. Rev. D \textbf{62}, 023511 (2000)

\bibitem{ArmendarizPicon:2000dh}
C.~Armendariz-Picon, V.F. Mukhanov, P.J. Steinhardt, Phys. Rev. Lett. \textbf{85}, 4438 (2000)

\bibitem{ArmendarizPicon:2000ah}
C.~Armendariz-Picon, V.F. Mukhanov, P.J. Steinhardt, Phys. Rev. D \textbf{63}, 103510 (2001)

\bibitem{2004PhRvD..70l3518B}
P.~{Brax}, C.~{van de Bruck}, A.C. {Davis}, J.~{Khoury}, A.~{Weltman}, prd \textbf{70}(12), 123518 (2004).
\newblock \doi{10.1103/PhysRevD.70.123518}

\bibitem{2004PhRvL..93q1104K}
J.~{Khoury}, A.~{Weltman}, prl \textbf{93}(17), 171104 (2004).
\newblock \doi{10.1103/PhysRevLett.93.171104}

\bibitem{2007PhRvD..76f4004H}
W.~{Hu}, I.~{Sawicki}, prd \textbf{76}(6), 064004 (2007).
\newblock \doi{10.1103/PhysRevD.76.064004}

\bibitem{PhysRevD.77.046009}
G.~Cognola, E.~Elizalde, S.~Nojiri, S.D. Odintsov, L.~Sebastiani, S.~Zerbini, Phys. Rev. D \textbf{77}, 046009 (2008).
\newblock \doi{10.1103/PhysRevD.77.046009}.
\newblock \urlprefix\url{https://link.aps.org/doi/10.1103/PhysRevD.77.046009}

\bibitem{Lopez:2021agu}
M.~L\'opez, G.~Otalora, N.~Videla, JCAP \textbf{10}, 021 (2021).
\newblock \doi{10.1088/1475-7516/2021/10/021}

\bibitem{Gonzalez-Espinoza:2021qnv}
M.~Gonzalez-Espinoza, R.~Herrera, G.~Otalora, J.~Saavedra, Eur. Phys. J. C \textbf{81}(8), 731 (2021).
\newblock \doi{10.1140/epjc/s10052-021-09542-6}

\bibitem{Gonzalez-Espinoza:2021mwr}
M.~Gonzalez-Espinoza, G.~Otalora, J.~Saavedra, JCAP \textbf{10}, 007 (2021).
\newblock \doi{10.1088/1475-7516/2021/10/007}

\bibitem{Gonzalez-Espinoza:2020jss}
M.~Gonzalez-Espinoza, G.~Otalora, Eur. Phys. J. C \textbf{81}(5), 480 (2021).
\newblock \doi{10.1140/epjc/s10052-021-09270-x}

\bibitem{Gonzalez-Espinoza:2020azh}
M.~Gonzalez-Espinoza, G.~Otalora, Phys. Lett. B \textbf{809}, 135696 (2020).
\newblock \doi{10.1016/j.physletb.2020.135696}

\bibitem{Gonzalez-Espinoza:2019ajd}
M.~Gonzalez-Espinoza, G.~Otalora, N.~Videla, J.~Saavedra, JCAP \textbf{08}, 029 (2019).
\newblock \doi{10.1088/1475-7516/2019/08/029}

\bibitem{Otalora:2014aoa}
G.~Otalora, Int. J. Mod. Phys. D \textbf{25}(02), 1650025 (2015).
\newblock \doi{10.1142/S0218271816500255}

\bibitem{Otalora:2013dsa}
G.~Otalora, Phys. Rev. D \textbf{88}, 063505 (2013).
\newblock \doi{10.1103/PhysRevD.88.063505}

\bibitem{Otalora:2013tba}
G.~Otalora, JCAP \textbf{1307}, 044 (2013)

\bibitem{Gonzalez-Espinoza:2023whd}
M.~Gonzalez-Espinoza, G.~Otalora, Y.~Leyva, J.~Saavedra, Eur. Phys. J. C \textbf{84}(3), 308 (2024).
\newblock \doi{10.1140/epjc/s10052-024-12653-5}

\bibitem{Nicolis:2008in}
A.~Nicolis, R.~Rattazzi, E.~Trincherini, Phys. Rev. D \textbf{79}, 064036 (2009)

\bibitem{Deffayet:2009wt}
C.~Deffayet, G.~Esposito-Farese, A.~Vikman, Phys. Rev. \textbf{D79}, 084003 (2009)

\bibitem{Baker:2017hug}
T.~Baker, E.~Bellini, P.G. Ferreira, M.~Lagos, J.~Noller, I.~Sawicki, Phys. Rev. Lett. \textbf{119}(25), 251301 (2017)

\bibitem{Sakstein:2017xjx}
J.~Sakstein, B.~Jain, Phys. Rev. Lett. \textbf{119}(25), 251303 (2017)

\bibitem{2012PhR...513....1C}
T.~{Clifton}, P.G. {Ferreira}, A.~{Padilla}, C.~{Skordis}, Physics Reports \textbf{513}(1), 1 (2012).
\newblock \doi{10.1016/j.physrep.2012.01.001}

\bibitem{Ostrogradsky:1850fid}
M.~Ostrogradsky, Mem. Acad. St. Petersbourg \textbf{6}(4), 385 (1850)

\bibitem{Horndeski:1974wa}
G.W. Horndeski, Int. J. Theor. Phys. \textbf{10}, 363 (1974)

\bibitem{Armendariz-Picon:2004say}
C.~Armendariz-Picon, JCAP \textbf{07}, 007 (2004).
\newblock \doi{10.1088/1475-7516/2004/07/007}

\bibitem{Koivisto:2008xf}
T.~Koivisto, D.F. Mota, JCAP \textbf{08}, 021 (2008).
\newblock \doi{10.1088/1475-7516/2008/08/021}

\bibitem{Gomez:2020sfz}
L.G. Gomez, Y.~Rodriguez, Phys. Dark Univ. \textbf{31}, 100759 (2021).
\newblock \doi{10.1016/j.dark.2020.100759}

\bibitem{Gonzalez-Espinoza:2022hui}
M.~Gonzalez-Espinoza, G.~Otalora, Y.~Leyva, J.~Saavedra, Eur. Phys. J. Plus \textbf{138}(7), 600 (2023).
\newblock \doi{10.1140/epjp/s13360-023-04237-1}

\bibitem{DeFelice:2016yws}
A.~De~Felice, L.~Heisenberg, R.~Kase, S.~Mukohyama, S.~Tsujikawa, Y.l. Zhang, JCAP \textbf{06}, 048 (2016)

\bibitem{Rodriguez-Benites:2023otm}
C.~Rodriguez-Benites, M.~Gonzalez-Espinoza, G.~Otalora, M.~Alva-Morales, Eur. Phys. J. C \textbf{84}(3), 276 (2024).
\newblock \doi{10.1140/epjc/s10052-024-12613-z}

\bibitem{DeFelice:2016uil}
A.~De~Felice, L.~Heisenberg, R.~Kase, S.~Mukohyama, S.~Tsujikawa, Y.l. Zhang, Phys. Rev. D \textbf{94}(4), 044024 (2016)

\bibitem{Nakamura:2019phn}
S.~Nakamura, R.~Kase, S.~Tsujikawa, JCAP \textbf{12}, 032 (2019)

\bibitem{DeFelice:2020icf}
A.~De~Felice, S.~Nakamura, S.~Tsujikawa, Phys. Rev. D \textbf{102}(6), 063531 (2020)

\bibitem{Cardona:2022lcz}
W.~Cardona, J.B. Orjuela-Quintana, C.A. Valenzuela-Toledo, JCAP \textbf{08}(08), 059 (2022)

\bibitem{Horndeski:1976gi}
G.W. Horndeski, J. Math. Phys. \textbf{17}, 1980 (1976)

\bibitem{Heisenberg:2014rta}
L.~Heisenberg, JCAP \textbf{05}, 015 (2014).
\newblock \doi{10.1088/1475-7516/2014/05/015}

\bibitem{Tasinato:2014eka}
G.~Tasinato, JHEP \textbf{04}, 067 (2014).
\newblock \doi{10.1007/JHEP04(2014)067}

\bibitem{Tasinato:2014mia}
G.~Tasinato, Class. Quant. Grav. \textbf{31}, 225004 (2014).
\newblock \doi{10.1088/0264-9381/31/22/225004}

\bibitem{Allys:2015sht}
E.~Allys, P.~Peter, Y.~Rodriguez, JCAP \textbf{02}, 004 (2016).
\newblock \doi{10.1088/1475-7516/2016/02/004}

\bibitem{BeltranJimenez:2016rff}
J.~Beltran~Jimenez, L.~Heisenberg, Phys. Lett. B \textbf{757}, 405 (2016).
\newblock \doi{10.1016/j.physletb.2016.04.017}

\bibitem{Heisenberg:2018acv}
L.~Heisenberg, JCAP \textbf{10}, 054 (2018).
\newblock \doi{10.1088/1475-7516/2018/10/054}

\bibitem{Heisenberg:2018mxx}
L.~Heisenberg, R.~Kase, S.~Tsujikawa, Phys. Rev. D \textbf{98}(2), 024038 (2018).
\newblock \doi{10.1103/PhysRevD.98.024038}

\bibitem{deRham:2014zqa}
C.~de~Rham, Living Rev. Rel. \textbf{17}, 7 (2014)

\bibitem{Will2014}
C.M. {Will}, Living Reviews in Relativity \textbf{17}(1), 4 (2014).
\newblock \doi{10.12942/lrr-2014-4}

\bibitem{will2018theory}
C.M. Will, \emph{Theory and experiment in gravitational physics} (Cambridge university press, 2018)

\bibitem{Hohmann:2017qje}
M.~Hohmann, A.~Sch\"arer, Phys. Rev. D \textbf{96}(10), 104026 (2017).
\newblock \doi{10.1103/PhysRevD.96.104026}

\bibitem{Gonzalez-Espinoza:2021nqd}
M.~Gonzalez-Espinoza, G.~Otalora, L.~Kraiselburd, S.~Landau, JCAP \textbf{05}(05), 010 (2022).
\newblock \doi{10.1088/1475-7516/2022/05/010}

\bibitem{Vainshtein:1972sx}
A.I. Vainshtein, Phys. Lett. B \textbf{39}, 393 (1972).
\newblock \doi{10.1016/0370-2693(72)90147-5}

\bibitem{Burrage:2010rs}
C.~Burrage, D.~Seery, JCAP \textbf{08}, 011 (2010).
\newblock \doi{10.1088/1475-7516/2010/08/011}

\bibitem{DeFelice:2011th}
A.~De~Felice, R.~Kase, S.~Tsujikawa, Phys. Rev. D \textbf{85}, 044059 (2012).
\newblock \doi{10.1103/PhysRevD.85.044059}

\bibitem{Kimura:2011dc}
R.~Kimura, T.~Kobayashi, K.~Yamamoto, Phys. Rev. D \textbf{85}, 024023 (2012).
\newblock \doi{10.1103/PhysRevD.85.024023}

\bibitem{Kase:2013uja}
R.~Kase, S.~Tsujikawa, JCAP \textbf{08}, 054 (2013).
\newblock \doi{10.1088/1475-7516/2013/08/054}

\bibitem{DeFelice:2016cri}
A.~De~Felice, L.~Heisenberg, R.~Kase, S.~Tsujikawa, Y.l. Zhang, G.B. Zhao, Phys. Rev. D \textbf{93}(10), 104016 (2016).
\newblock \doi{10.1103/PhysRevD.93.104016}

\bibitem{2003Natur.425..374B}
B.~{Bertotti}, L.~{Iess}, P.~{Tortora}, nat \textbf{425}(6956), 374 (2003).
\newblock \doi{10.1038/nature01997}

\bibitem{LIGOScientific:2017vwq}
B.P. Abbott, et~al., Phys. Rev. Lett. \textbf{119}(16), 161101 (2017).
\newblock \doi{10.1103/PhysRevLett.119.161101}

\bibitem{Goldstein:2017mmi}
A.~Goldstein, et~al., Astrophys. J. Lett. \textbf{848}(2), L14 (2017).
\newblock \doi{10.3847/2041-8213/aa8f41}

\bibitem{Copeland:1997et}
E.J. Copeland, A.R. Liddle, D.~Wands, Phys. Rev. D \textbf{57}, 4686 (1998).
\newblock \doi{10.1103/PhysRevD.57.4686}

\bibitem{coefPPN}
\url{https://github.com/manuelgonzalez-upla/Screening-fifth-forces-in-scalar-vector-tensor-theories}

\end{thebibliography}

\appendix
\section{Appendix: $\mathcal{C}_i$ and $\mathcal{D}_i$ }\label{Appendix0}
In this appendix we show coefficients $\mathcal{C}_i$ and $\mathcal{D}_i$ 
\begin{eqnarray}
    \mathcal{C}_1 &=& 4 e^{2 \Psi } \phi _A^2 f_{2, F}, \\
    \mathcal{C}_2 &=& 4 e^{2 \Psi } \phi _A^2 \chi ' f_{3, X_3} + 4 e^{2 \Psi } \phi _A \phi _A' f_{2, F}, \\
    \mathcal{C}_3 &=& 2 e^{-2 \Phi } \left(\chi '\right)^3 f_{3, X_3}-2 e^{2 \Psi } \phi _A^2 \chi ' f_{3,X_3}-2 \phi ' f_{4,\phi}, \\
    \mathcal{C}_4 &=& -4 f_{4,\phi}, \\
    \mathcal{C}_5 &=& \phi _A^2 e^{2 \Phi +2 \Psi } f_{2,X_3}-e^{2 \Phi } f_2+e^{2 \Psi } \left(\phi _A'\right){}^2 f_{2, F}+2 \chi ' \phi ' f_{3, \phi}+2 e^{2 \Psi } \phi _A \chi ' \phi _A' f_{3, X_3}\nonumber\\
    &-& 2 e^{-2 \Phi } \left(\chi '\right)^2 \chi '' f_{3, X_3} +2 f_{4 ,\phi} \phi ''(r)+2 \left(\phi '\right)^2 f_{4, \phi \phi } +2 e^{2 \Psi } \phi _A^2 \chi '' f_{3, X_3}, \\
    \mathcal{C}_6 &=& 4 e^{2 \Psi } \phi _A^2 \chi ' f_{3, X_3}+4 \phi ' f_{4, \phi}, \\ 
    \mathcal{C}_7 &=& 2 f_4-2 e^{2 \Phi } f_4, \\
    \mathcal{C}_8 &=& 4 e^{2 \Psi } \phi _A^2 f_{2, F}, \\
    \mathcal{C}_9 &=& 4 e^{2 \Psi } \phi _A \phi _A' f_{2, F}-2 e^{-2 \Phi } \left(\chi '\right)^3 f_{3, X_3}-2 e^{2 \Psi } \phi _A^2 \chi ' f_{3, X_3}+2 \phi ' f_{4, \phi}, \\
    \mathcal{C}_{10} &=& 4 f_4, \\
    \mathcal{C}_{11} &=& \left(\phi '\right)^2 \left(-f_{2,X_1}\right)-\chi ' \phi ' f_{2,X_2}-e^{2 \Phi } f_2+e^{2 \Psi } \left(\phi _A'\right){}^2 f_{2,F}-\left(\chi '\right)^2 f_{2,X_3}-2 \chi ' \phi ' f_{3, X_3} \nonumber\\
    &-&2 e^{2 \Psi } \phi _A \chi ' \phi _A' f_{3, X_3}, \\
    \mathcal{C}_{12} &=& 4 \phi ' f_{4, \phi}-4 e^{-2 \Phi } \left(\chi '\right)^3 f_{3, X_3}, \\
    \mathcal{C}_{13} &=& 2 f_4-2 e^{2 \Phi } f_4. \\
    \mathcal{C}_{14} &=& 2 f_4, \\
    \mathcal{C}_{15} &=& 2 f_4, \\
    \mathcal{C}_{16} &=& -2 f_4, \\
    \mathcal{C}_{17} &=& 2 e^{2 \Psi } f_{3,X_3} \phi _A^2 \chi '-2 f_4 \Phi '+2 \phi ' f_{4,\phi}, \\
    \mathcal{C}_{18} &=& 2 f_4, \\
    \mathcal{C}_{19} &=& 2 e^{2 \Psi } f_{3, X_3} \phi _A \chi ' \phi _A'-2 e^{-2 \Phi } f_{3, X_3} \left(\chi '\right)^2 \chi ''+2 f_{3, \phi} \chi ' \phi '+2 f_{4,\phi} \phi ''(r)-f_2 e^{2 \Phi } \nonumber\\
    &&+2 \left(\phi '\right)^2 f_{4,\phi \phi}, \\ 
    \mathcal{C}_{20} &=& 2 f_{4, \phi} \phi ',\\ 
     \mathcal{D}_1 &=& 4 e^{2 \Psi } f_{2, X_1 F} \phi _A^2 \Psi ' \phi '+2 e^{2 \Psi } f_2{2, X_1 F} \phi _A \phi ' \phi _A'+2 e^{2 \Psi } f_{2, X_2 F} \phi _A^2 \chi ' \Psi '+e^{2 \Psi } f_{2, X_2 F} \phi _A \chi ' \phi _A' \nonumber\\
    &&-2 e^{2 \Phi } f_4'(\phi ), \\
    \mathcal{D}_2 &=& 4 e^{2 \Psi } f_{2, X_1 F} \phi _A^2 \phi '+2 e^{2 \Psi } f_{2, X_2 F} \phi _A^2 \chi ', \\
    \mathcal{D}_3 &=& -4 e^{2 \Psi } f_{2, X_1 F} \phi _A^2 \phi ' \Phi '-2 e^{2 \Psi } f_{2, X_2 F} \phi _A^2 \chi ' \Phi '+8 e^{2 \Psi } f_{2, X_1 F} \phi _A \phi ' \phi _A'+4 e^{2 \Psi } f_{2, X_2 F} \phi _A \chi ' \phi _A' \nonumber\\
    &&-2 e^{2 \Phi } f_4'(\phi ), \\
    \mathcal{D}_4 &=& 4 e^{2 \Psi } f_{2, X_1 F} \phi _A^2 \phi '+2 e^{2 \Psi } f_{2, X_2 F} \phi _A^2 \chi ', \\
    \mathcal{D}_5 &=& -4 e^{2 \Psi } f_{2, X_1 F} \phi _A^2 \phi '-2 e^{2 \Psi } f_{2, X_2 F} \phi _A^2 \chi ', \\
    \mathcal{D}_6 &=& -4 e^{2 \Psi } f_{2, X_1 F} \phi _A \Phi ' \phi ' \phi _A'+f_{2, X_1 X_3} \phi _A^2 e^{2 \Phi +2 \Psi } \phi '+4 e^{2 \Psi } f_{2, X_1 F} \phi _A^2 \Psi '' \phi '+2 e^{2 \Psi } f_{2, X_1 F} \phi _A \phi ' \phi _A''\nonumber\\
    &&+3 e^{2 \Psi } f_{2, X_1 F} \phi ' \left(\phi _A'\right){}^2-2 e^{2 \Psi } f_{2, X_2 F} \phi _A \Phi ' \chi ' \phi _A'+\frac{1}{2} f_{2, X_2 X_3} \phi _A^2 e^{2 \Phi +2 \Psi } \chi '+2 e^{2 \Psi } f_{2, X_2 F} \phi _A^2 \chi ' \Psi ''\nonumber\\
    &&+e^{2 \Psi } f_{2, X_2 F} \phi _A \chi ' \phi _A''+\frac{3}{2} e^{2 \Psi } f_{2, X_2 F} \chi ' \left(\phi _A'\right){}^2+2 e^{2 \Phi } f_{2, \phi} \chi '+\frac{1}{2} e^{2 \Phi } f_{2, X_2} \chi '+e^{2 \Phi } f_{2, X_1} \phi '\nonumber\\
    &&+2 e^{2 \Phi } \Phi ' f_{4,\phi},  
\end{eqnarray}
\begin{eqnarray}
    \mathcal{D}_7 &=& -4 e^{2 \Phi } f_{4,\phi}, \\
    \mathcal{D}_8 &=& -4 e^{2 \Psi } f_{2, X_1 F} \phi _A^2 \left(\Psi '\right)^2 \phi '-4 e^{2 \Psi } f_{2, X_1 F} \phi _A \Psi ' \phi ' \phi _A'-e^{2 \Psi } f_{2, X_1 F} \phi ' \left(\phi _A'\right){}^2-2 e^{2 \Psi } f_{2, X_2 F} \phi _A^2 \chi ' \left(\Psi '\right)^2\nonumber\\
    &&-2 e^{2 \Psi } f_{2, X_2 F} \phi _A \chi ' \Psi ' \phi _A'-\frac{1}{2} e^{2 \Psi } f_{2, X_2 F} \chi ' \left(\phi _A'\right){}^2-2 e^{2 \Phi } f_{3, \phi} \chi '-\frac{1}{2} e^{2 \Phi } f_{2, X_2} \chi '+\frac{1}{2} f_{2, X_2 X_3} \left(\chi '\right)^3\nonumber\\
    &&-e^{2 \Phi } f_{2, X_1} \phi '+\frac{3}{2} f_{2, X_1 X_2} \chi ' \left(\phi '\right)^2+\frac{1}{2} f_{2, X_2 X_2} \left(\chi '\right)^2 \phi '+f_{2, X_2 X_3} \left(\chi '\right)^2 \phi '+f_{2, X_1 X_1} \left(\phi '\right)^3\nonumber\\
    &&+2 e^{2 \Phi } \Psi ' f_{4_\phi }, \\
    \mathcal{D}_9 &=& 4 e^{2 \Phi } f_{4_\phi }, \\
    \mathcal{D}_{10} &=& -4 e^{2 \Psi } f_{2, X_1 F} \phi _A \phi ' \phi _A'-2 e^{2 \Psi } f_{2, X_2 F} \phi _A \chi ' \phi _A'+2 e^{2 \Phi } f_{4_\phi }, \\
    \mathcal{D}_{11} &=& f_{2, X_1 X_3} \phi _A e^{2 \Phi +2 \Psi } \phi ' \phi _A'+e^{2 \Psi } f_{2, X_1 F} \phi ' \phi _A' \phi _A''+\frac{1}{2} f_{2, X_2 X_3} \phi _A e^{2 \Phi +2 \Psi } \chi ' \phi _A'+\frac{1}{2} e^{2 \Psi } f_{2, X_2 F} \chi ' \phi _A' \phi _A''\nonumber\\
    &&+e^{2 \Phi } f_{2, X_1} \phi ''(r)-\frac{1}{4} f_{2, X_2 X_2} \left(\chi '\right)^2 \phi ''(r)-f_{2, X_1 X_2} \chi ' \phi ' \phi ''(r)-f_{2, X_1 X_1} \left(\phi '\right)^2 \phi ''(r)+2 e^{2 \Phi } f_{3, \phi} \chi ''\nonumber\\
    &&+\frac{1}{2} e^{2 \Phi } f_{2, X_2} \chi ''+e^{4 \Phi } f_{2, \phi}-\frac{1}{2} f_{2, X_2 X_3} \left(\chi '\right)^2 \chi ''+\frac{1}{2} e^{2 \Phi } f_{2, \phi X_2} \chi ' \phi '+e^{2 \Phi } f_{2,\phi X_1} \left(\phi '\right)^2\nonumber\\
    &&-\frac{1}{2} f_{2, X_1 X_2} \chi '' \left(\phi '\right)^2-\frac{1}{4} f_{2, X_2 X_2} \chi ' \chi '' \phi '-f_{2, X_1 X_3} \chi ' \chi '' \phi ', \\
    \mathcal{D}_{12} &=& 4 e^{2 \Phi } f_{3, \phi} \chi '+e^{2 \Phi } f_{2, X_2} \chi '+2 e^{2 \Phi } f_{2, \phi X_1} \phi ', \\
    \mathcal{D}_{13} &=& 4 e^{2 \Phi } f_{3, \phi} \chi '+e^{2 \Phi } f_{2, X_2} \chi '+2 e^{2 \Phi } f_{2, \phi X_1} \phi ' \\
    \mathcal{D}_{14} &=& 2 e^{2 \Phi } f_{2, F} \phi _A+8 e^{2 \Psi } f_{2, F F} \phi _A^3 \left(\Psi '\right)^2+8 e^{2 \Psi } f_{2, F F} \phi _A^2 \Psi ' \phi _A'+2 e^{2 \Psi } f_{2, F F} \phi _A \left(\phi _A'\right){}^2, \\
    \mathcal{D}_{15} &=& 8 e^{2 \Psi } f_{2, F F} \phi _A^2 \phi _A', \\
    \mathcal{D}_{16} &=& 8 e^{2 \Psi } f_{2, F F} \phi _A^3, \\
    \mathcal{D}_{17} &=& -12 e^{2 \Psi } f_{2, F F} \phi _A^2 \Phi ' \phi _A'+2 f_{2, X_3 F} \phi _A^3 e^{2 \Phi +2 \Psi }+2 e^{2 \Phi } f_{2, F} \phi _A+8 e^{2 \Psi } f_{2, F F} \phi _A^3 \Psi ''\nonumber\\
    &&+14 e^{2 \Psi } f_{2, F F} \phi _A \left(\phi _A'\right){}^2+4 e^{2 \Psi } f_{2, F F} \phi _A^2 \phi _A'', \\
    \mathcal{D}_{18} &=& 20 e^{2 \Psi } f_{2, F F} \phi _A^2 \phi _A'-8 e^{2 \Psi } f_{2, F F} \phi _A^3 \Phi ', \\
    \mathcal{D}_{19} &=& 8 e^{2 \Psi } f_{2, F F} \phi _A^3, \\ 
    \mathcal{D}_{20} &=& -12 e^{2 \Psi } f_{2, F F} \phi _A^2 \phi _A', \\
    \mathcal{D}_{21} &=& -8 e^{2 \Psi } f_{2, F F} \phi _A^3, \\
    \mathcal{D}_{22} &=& -f_{2, X_2 F} \phi _A \chi ' \phi ''-2 f_{2, X_1 F} \phi _A \phi ' \phi ''+2 f_{2, X_2 F} \phi _A \Phi ' \chi ' \phi '+2 f_{2, X_1 F} \phi _A \Phi ' \left(\phi '\right)^2+2 e^{2 \Phi } f_{2, \phi F} \phi _A \phi '\nonumber\\
    &&-f_{2, X_2 F} \phi _A \chi '' \phi '+2 f_{2, X_3 F} \phi _A \Phi ' \left(\chi '\right)^2-6 e^{2 \Psi } f_{2, F F} \phi _A \Phi ' \left(\phi _A'\right){}^2-2 e^{2 \Phi } f_{2, F} \phi _A \Phi '-2 e^{2 \Phi } f_3{}^{(0,1)} \phi _A \chi '\nonumber\\
    &&+3 f_{2, X_3 F} \phi _A^2 e^{2 \Phi +2 \Psi } \phi _A'+3 e^{2 \Phi } f_{2, F} \phi _A'-2 f_{2, X_3 F} \phi _A \chi ' \chi ''+8 e^{2 \Psi } f_{2, F F} \phi _A^2 \Psi '' \phi _A'+4 e^{2 \Psi } f_{2, F F} \phi _A \phi _A' \phi _A''\nonumber\\
    &&+3 e^{2 \Psi } f_{2, F F} \left(\phi _A'\right){}^3, \\
    \mathcal{D}_{23} &=& 4 e^{2 \Phi } f_{2, F} \phi _A, \\
    \mathcal{D}_{24} &=& 2 f_{2, X_2 F} \phi _A \chi ' \Psi ' \phi '+f_{2, X_2 F} \chi ' \phi ' \phi _A'+2 f_{2, X_1 F} \phi _A \Psi ' \left(\phi '\right)^2+f_{2, X_1 F} \left(\phi '\right)^2 \phi _A'+2 e^{2 \Phi } f_{3, X_3} \phi _A \chi '\nonumber\\
    &&-2 e^{2 \Phi } f_{2, F} \phi _A \Psi '-e^{2 \Phi } f_{2, F} \phi _A'+2 f_{2, X_3 F} \phi _A \left(\chi '\right)^2 \Psi '+f_{2, X_3 F} \left(\chi '\right)^2 \phi _A'-8 e^{2 \Psi } f_{2, F F} \phi _A^3 \left(\Psi '\right)^3\nonumber\\
    &&-12 e^{2 \Psi } f_{2, F F} \phi _A^2 \left(\Psi '\right)^2 \phi _A'-6 e^{2 \Psi } f_{2, F F} \phi _A \Psi ' \left(\phi _A'\right){}^2-e^{2 \Psi } f_{2, F F} \left(\phi _A'\right){}^3 \\
    \mathcal{D}_{25} &=& 2 f_{2, X_2 F} \phi _A \chi ' \phi '+2 f_{2, X_1 F} \phi _A \left(\phi '\right)^2-2 e^{2 \Phi } f_{2, F} \phi _A+2 f_{2, X_3 F} \phi _A \left(\chi '\right)^2-6 e^{2 \Psi } f_{2, F F} \phi _A \left(\phi _A'\right){}^2, \\
    \mathcal{D}_{26} &=& -\frac{1}{2} f_{2, X_2 F} \chi ' \phi _A' \phi ''-f_{2, X_1 F} \phi ' \phi _A' \phi ''+e^{2 \Phi } f_{2, \phi F} \phi ' \phi _A'-\frac{1}{2} f_{2, X_2 F} \chi '' \phi ' \phi _A'-2 e^{2 \Phi } f_{3, X_3} \phi _A \chi ''\nonumber\\
    &&+f_{2, X_3 F} \phi _A e^{2 \Phi +2 \Psi } \left(\phi _A'\right){}^2+e^{2 \Phi } f_{2, F} \phi _A''-e^{4 \Phi } f_{2, X_3} \phi _A-f_{2, X_3 F} \chi ' \chi '' \phi _A'+e^{2 \Psi } f_{2, F F} \left(\phi _A'\right){}^2 \phi _A'', \\
    \mathcal{D}_{27} &=& 2 e^{2 \Phi } f_{2, F} \phi _A'-4 e^{2 \Phi } f_{3, X_3} \phi _A \chi ', \\
    \mathcal{D}_{28} &=& 2 e^{2 \Phi } f_{2, F} \phi _A'-4 e^{2 \Phi } f_{3, X_3} \phi _A \chi ', \\
    \mathcal{D}_{29} &=& -4 e^{2 \Psi } f_{3, X_3} \phi _A^2-4 e^{-2 \Phi } f_{3, X_3} \left(\chi '\right)^2, \\
    \mathcal{D}_{30} &=& -4 e^{2 \Psi } f_{3, X_3} \phi _A \phi _A'-2 f_{2, X_3} \chi '-4 f_{3, \phi} \phi '-f_{2, X_2} \phi ', \\ 
    \mathcal{D}_{31} &=& -8 e^{-2 \Phi } f_{3, X_3} \left(\chi '\right)^2, \\
    \mathcal{D}_{32} &=& -8 e^{-2 \Phi } f_{3, X_3} \left(\chi '\right)^2 . 
\end{eqnarray}

\section{Appendix: General Relativity} \label{Appendix1}
In GR, we have
$f_2=f_3=0$ and $f_4=M_{\rm pl}^2/2$ , so equations (\ref{eom00}) and (\ref{eom11}) read
\begin{eqnarray}
&&
\frac{2M_{\rm pl}^2}{r}\Phi_{\rm GR}'
-\frac{M_{\rm pl}^2}{r^2}
\left( 1-e^{2\Phi_{\rm GR}} \right)
=e^{2\Phi_{\rm GR}} \rho_m\,,
\label{GR1}\\
&&
\frac{2M_{\rm pl}^2}{r}\Psi_{\rm GR}'
+\frac{M_{\rm pl}^2}{r^2}
\left( 1-e^{2\Phi_{\rm GR}} \right)
=e^{2\Phi_{\rm GR}} P_m\,.
\label{GR2}
\end{eqnarray}

$\rho_m(r) \simeq \rho_0$ for $r<r_*$ and $\rho_m(r) \simeq 0$ for $r>r_*$, being $M_{\rm pl}$ the reduced Planck mass.
The gravitational potentials inside and outside the body are given by
\be
e^{\Psi_{\rm GR}}=\frac32 \sqrt{1-\frac{\rho_0 r_*^2}
{3M_{\rm pl}^2}}
-\frac12 \sqrt{1-\frac{\rho_0 r^2}{3M_{\rm pl}^2}}\,,\qquad
e^{\Phi_{\rm GR}}= \left( 1-\frac{\rho_0 r^2}{3M_{\rm pl}^2}\right)^{-1/2}\,,
\label{Sin}
\ee
for $r<r_*$, and
\be
e^{\Psi_{\rm GR}}=\left( 1 -\frac{\rho_0 r_*^3}
{3M_{\rm pl}^2 r} \right)^{1/2}\,,
\qquad
e^{\Phi_{\rm GR}}=\left( 1 -\frac{\rho_0 r_*^3}
{3M_{\rm pl}^2 r} \right)^{-1/2}\,,
\label{Sout}
\ee
for $r>r_*$. In the following, we employ the weak
gravity approximation under which $|\Psi|$ and $|\Phi|$ are much smaller than 1, i.e.,
\be
\Phi_0 \equiv
\frac{\rho_0 r_*^2}{M_{\rm pl}^2} \ll 1\,.
\label{weakcon}
\ee
This condition means that the Schwarzschild radius
of the source $r_g \approx \rho_0 r_*^3/M_{\rm pl}^2$
is much smaller than $r_*$.
Then, the solutions (\ref{Sin}) and (\ref{Sout}) reduce,
respectively, to
%
%
\begin{eqnarray}
    \Psi_{GR} \simeq \dfrac{\rho_0}{12 M_{\text{pl}}^2} (r^2 - 3 r^2_*),& \ \ \ \ \ \ \ \Phi_{GR} \simeq \dfrac{\rho_0 r^2}{6 M_{\text{pl}}^2}, & \ \ \ \ \ \ \ \text{for} \ r<r_* ,\label{GR_potential_inside} \\
    \Psi_{GR} \simeq -\dfrac{\rho_0 r^2}{6 M_{\text{pl}}^2},& \ \ \ \ \ \ \ \Phi_{GR} \simeq \dfrac{\rho_0 r^2}{6 M_{\text{pl}}^2}, & \ \ \ \ \ \ \ \text{for} \ r>r_* . \label{GR_potential_outside}
\end{eqnarray} 

\end{document}